\documentclass[aps,prd,twocolumn,nofootinbib,superscriptaddress]{revtex4-2}
\pdfoutput=1
\usepackage{graphicx}  
\usepackage{dcolumn}   
\usepackage{bm}        
\usepackage{amsmath,amssymb,amsxtra,bm,epsfig}   
\usepackage{float}
\usepackage[dvipsnames]{xcolor}
\usepackage{xspace}
\usepackage{natbib}
\usepackage{hyperref}
\hypersetup{
	colorlinks=true,
	linkcolor=red,
	filecolor=magenta,
	urlcolor=ForestGreen,
	citecolor=RoyalBlue}

\usepackage[]{caption}
\begin{document}
	\title{Flavor Leptogenesis During Reheating Era}
	
%
%
%
%
	
		\author{Arghyajit Datta}
		\email{datta176121017@iitg.ac.in}
		\affiliation{Department of Physics, Indian Institute of Technology Guwahati, Assam 781039, India}
	
	\author{Rishav Roshan}
	\email{rishav.roshan@gmail.com}
	\affiliation{Department of Physics, Kyungpook National University, Daegu 41566, Korea}
	
	\author{Arunansu Sil}
	\email{asil@iitg.ac.in}
	\affiliation{Department of Physics, Indian Institute of Technology Guwahati, Assam 781039, India}
		\begin{abstract}
		Recently, it has been shown that the presence of a non-instantaneous era of reheating can significantly alter the charged lepton(s) equilibration temperature(s) which plays important role in flavor leptogenesis. In this work, we extend the analysis to a more general situation where RHNs are also produced from the decay of the inflaton. The presence of these RHNs along with the thermally generated ones (above its mass equivalent temperature only) redistributes different components of the energy density of the Universe during this reheating era, thereby affecting the charged lepton equilibration temperature (in addition to the Hubble effect) as well as the final reheating temperature $T_{\rm{RH}}$. Taking both the effects into account, we find that the decay of the lightest RHN in the set-up not only provides a platform to study flavor leptogenesis during reheating, but also an interesting framework of $quasi$-thermal leptogenesis emerges. 		
		\end{abstract}
	%
	\maketitle

\section{Introduction}

The existence of non-zero neutrino masses~\cite{Fukuda:1998mi,Ahmad:2002jz,Ahn:2002up}, and the origin of baryon asymmetry~\cite{Planck:2018vyg} of the Universe (BAU) are two of the major issues that the Standard Model (SM) of particle physics fails to accommodate, indicating the necessity for the physics beyond the SM. While the issue of light neutrino mass can be elegantly handled by the introduction of three heavy SM singlet right-handed neutrinos (RHN) having Yukawa interaction with the SM Higgs and lepton doublets within the so called `seesaw' mechanism~\cite{Minkowski:1977sc,Yanagida:1979as,Yanagida:1979gs,GellMann:1980vs,Mohapatra:1979ia,Schechter:1980gr,Schechter:1981cv,Datta:2021elq}, the same offers an interesting explanation of the BAU through leptogenesis \cite{Fukugita:1986hr,Luty:1992un,Pilaftsis:1997jf}. Here, a lepton asymmetry is created as a result of the CP-violating out-of-equilibrium decay of the RHNs which then be partially converted into the baryon asymmetry through the $(B+L)$ violating sphaleron interactions of the SM at electroweak temperature $T_{\rm EW} \gtrsim$ 100 GeV. 

In most widely studied framework of `$thermal$' leptogenesis, it is considered that after the Universe enters in the radiation dominated era (the beginning of which is marked by reheating temperature $T_{\text{RH}}$), the RHNs can be created by thermal scattering more specifically via inverse decays and 2-2 scattering mediated by SM fields. Subsequently, considering a hierarchical RHN masses, the lightest among the three RHNs (say $N_1$ with mass $M_1$) starts to contribute to lepton asymmetry production via its out-of-equilibrium decay to the SM lepton ($l_{L_{\alpha}}$) and Higgs ($H$) doublets around a temperature $ T \lesssim M_1$. The reheating temperature obviously satisfies the condition $T_{\text{RH}} > M_1$. The abundance of the RHNs ($Y_{N_1}$) and the produced lepton asymmetry in a specific flavor direction ($\Delta_{L_{\alpha}}$) in this radiation dominated epoch are connected by the Boltzmann equations (BE) where apart from production, one needs to incorporate all the lepton-number violating processes that can potentially erase such asymmetry.  

Provided that such decay happens at sufficiently high temperature (at or above $5\times10^{11}$ GeV), where all the right-handed charged leptons of different flavors are in out of equilibrium, one can safely use the unflavored approximation \cite{Fukugita:1986hr,Luty:1992un,Pilaftsis:1997jf}. This is because the rate of charged lepton Yukawa interactions remain weaker compared to that of RHN Yukawa interactions. On the other hand, if the leptogenesis happens at a temperature ($T$) below $5\times10^{11}$ GeV, the right-handed tau leptons equilibrate with the thermal bath and flavor effects ~\cite{Barbieri:1999ma,Nardi:2005hs,Nardi:2006fx,Abada:2006fw,Abada:2006ea,Blanchet:2006ch,Blanchet:2006be,Dev:2017trv,Datta:2021gyi} are inevitable. Once this tau lepton Yukawa interaction is equilibrated, it tends to destroy the lepton asymmetry carried by the tau leptons generated from the decay of RHNs. For the muon and electron Yukawa interactions, this happens below $10^{9}$ GeV and $5\times10^{4}$ GeV respectively. 

There exists a lower limit on the mass of the RHNs as $M_1 \gtrsim 10^9$ GeV (known as Davidson-Ibarra bound \cite{Davidson:2002qv}) in order to satisfy the correct baryon asymmetry of the Universe via leptogenesis which in turn indicates that reheating temperature should be higher than this value for standard {\it thermal} leptogenesis. Although it is feasible to have such a high $T_{\text{RH}}$, there is no such concrete evidence in support of it too. On the contrary, it can be as low as few MeV~\cite{Giudice:2000ex,Kawasaki:2000en,Martin:2010kz,Dai:2014jja}. In this context, it is interesting to investigate the possibility of having reheating temperature smaller than the mass of the RHNs in view of leptogenesis. While one such possibility is to have {\it non-thermal} leptogenesis \cite{Lazarides:1991wu,Murayama:1992ua,Kolb:1996jt,Giudice:1999fb,Asaka:1999yd,Asaka:1999jb,Hamaguchi:2001gw,Jeannerot:2001qu,Fujii:2002jw,Giudice:2003jh,Pascoli:2003rq,Asaka:2002zu,Panotopoulos:2006wj,HahnWoernle:2008pq,Hamada:2015xva,Borah:2020wyc,Samanta:2020gdw,Barman:2021tgt,Azatov:2021irb,Barman:2021ost,Barman:2022gjo,Lazarides:2022spe,Lazarides:2022ezc,Ghoshal:2022fud,Ghoshal:2022kqp}, a different possibility opens up where a non-instantaneous reheating period (extended from $T_{\text{Max}}$ to $T_{\text{RH}}$) can be brought into the picture. It is known that reheating can actually be a non-instantaneous process \cite{Chung:1998rq,Giudice:2000ex,Giudice:2000ex,Mukaida:2015ria,Harigaya:2019tzu,Garcia:2020eof,Haque:2020zco} where a maximum temperature $T_\text{Max}$ after inflation can be realized followed by the onset of radiation dominated era indicated by reheating temperature $T_{\text{RH}}$ with 
$T_{\text{Max}} > T_{\text{RH}}$. In a recent study \cite{Datta:2022jic}, we have shown that leptogenesis remains a viable option even if the decaying RHN mass (lightest and hence responsible for lepton asymmetry generation) satisfies $T_{\text{Max}} > M_1 > T_{\text{RH}}$. 

Additionally in ~\cite{Datta:2022jic}, we have an important observation (for the first time) that a prolonged reheating period modifies the equilibration temperature (ET) of individual charged lepton Yukawa interactions and hence the study of leptogenesis, in particular the flavor leptogenesis during this extended reheating period becomes very rich. Here, the effective coupling of the inflaton with the SM fermion fields provides the sole contribution to the radiation component of the Universe defining the temperature of the thermal bath. This, in turn, plays a non-trivial role in controlling the expansion rate of the Universe during the course of reheating. A not-so-small choice of the effective coupling leads to a faster expansion of the Universe in the reheating period which carries the potential to delaying the era of equilibration of different charged lepton Yukawa interactions. Once the RHNs are thermally produced from the bath, and subsequently decay out of equilibrium (beyond $T \lesssim M_1$) in the period of this extended reheating, it is found that the delayed equilibration of the charged lepton Yukawa interaction can significantly shift the flavor regimes of the leptogenesis.

In this work, we focus on a more general picture allowing an additional interaction between the inflaton and the RHN fields
on top of the existing effective coupling between the inflaton and SM fermion fields, designated by $y_{\phi ff}$ coupling,  as in \cite{Datta:2022jic}. Hence, apart from thermal scattering, RHNs ($N_1$ here for simplicity) may also be produced directly from the decay of the inflaton field. This new source of RHN production can in principle alters, depending on the relative strength of the RHN-inflaton interaction (indicated by $y_{\phi NN}$), the individual components (such as for inflaton, RHNs and radiation) of energy density of the Universe during the extended reheating period. Then, a subsequent effect on the Hubble expansion not only affects the final reheating temperature but also modifies the ET of individual charged lepton flavor with respect to the one observed in \cite{Datta:2022jic}. 

Furthermore, for the lightest RHN mass lies in between $T_{\text{Max}}$ and $T_{\text{RH}}$, we encounter an interesting situation for leptogenesis here. We have already found that a modified $thermal$ (flavored) leptogenesis results in this extended period of reheating due to the absence of radiation domination as well as shift in the charged lepton ET. This observation can be visualized as a limiting case of vanishing inflaton-RHN coupling of the present proposal. In this general set-up, once this inflaton-RHN coupling $y_{\phi NN}$ is switched on, injection of the non-thermal RHNs into the system on top of the thermally generated ones is expected to enhance the outcome of leptogenesis. However, with not-so-large inflaton-RHN coupling, the result remains essentially close to {\it thermal} leptogenesis in extended reheating scenario. However, once $y_{\phi NN}$ becomes significant enough, say comparable to or larger than $y_{\phi ff}$, the RHNs produced from inflaton decay along with thermally generated ones could stay out-of-equilibrium during this reheating period itself. As a result, these RHNs may  effectively decay above the temperature $T\sim M_1$ and start to produce lepton asymmetry at $T > M_1$. Therefore, for such moderate range of $y_{\phi NN}$ coupling, we realize a situation which is intermediate between purely {\it thermal} and {\it non-thermal} leptogenesis scenario, which we name as {\it quasi}-thermal leptogenesis. Note that it does not indicate a completely new direction for leptogenesis, rather it corresponds to a more general and detailed interplay of thermal and non-thermal contributions of RHN production toward leptogenesis during an extended reheating period inclusive of flavor effect (that affects differently to thermal and non-thermal contributions) in comparison to earlier studies \cite{Hahn-Woernle:2008tsk,Buchmuller:2012wn,Buchmuller:2013dja}.  It is found that for sufficiently large $y_{\phi NN}$, the presence of accumulated number density of RHNs helps relaxing the lower limit of the RHN mass $M_1$ to some extent for which adequate lepton asymmetry can be produced.

The paper is organized as follows. Below in section~\ref{section:2}, we provide a brief overview of the standard $thermal$ leptogenesis and importance of flavors while in section \ref{section:3}, we discuss our general set-up of {\it quasi}-thermal leptogenesis. We devote section \ref{section:4} to discuss the outcome of the proposal. Finally in section \ref{section:5}, we conclude.

\section{Thermal Leptogenesis and effect of flavor}
\label{section:2}

A mere extension of the SM by three right handed singlet neutrinos ($N_{i=1,2, 3}$) as suggested by the type-I seesaw forms the basic set-up to discuss leptogenesis, the Lagrangian of which (in the charged lepton diagonal basis) is given by 
\begin{align}
	-\mathcal{L}_{T_I}= \overline{\ell}_{L_\alpha} (Y_{\nu})_{\alpha i} \tilde{H} N_{i}+ \frac{1}{2}  \overline{N_{i}^c}(M_{R})_{ii} N_i+ h.c.,
	\label{eq:1}
\end{align}
where the lepton number violating Majorana mass term for RHNs, $M_R$, is considered to be diagonal, $M_R = {\rm{diag}}(M_1, M_2, M_3)$ for simplicity. The neutrino Yukawa coupling $Y_{\nu}$ matrix in general contains CP violating phases. A Dirac mass term $m_D= Y_{\nu}v/\sqrt{2}$ is generated 
after spontaneous breaking of electroweak symmetry with $v$ = 246 GeV. In the see-saw limit $m_D \ll M_R$, 
a light neutrino mass matrix 
\begin{align}
	m_\nu=-m_D M_R^{-1} m_D^T,
\end{align}
results along with three heavy neutrinos. A further diagonalization of $m_{\nu}$ by the PMNS matrix $U$ \cite{Esfahani:2017dmu,Esteban:2020cvm,Zyla:2020zbs} via $U^{\dagger} m_{\nu} U^*$= diag ($m_1, m_2, m_3$), leads to three light neutrino masses $m_{i=1,2,3}$. 

The same seesaw Lagrangian also provides a natural explanation of the matter anti-matter asymmetry of the Universe via leptogenesis \cite{Fukugita:1986hr,Luty:1992un,Pilaftsis:1997jf} where the CP-violating decays of heavy RHNs into SM lepton and Higgs doublets, $N_i\to \ell_{L_\alpha} +H$, are instrumental. In the early Universe, provided the temperature (after inflation) was high enough, these heavy RHNs can be produced from thermal bath via inverse decay (mediated by the same neutrino Yukawa interaction) and attain thermal equilibrium. Thereafter, as the temperature drops below the individual mass of a RHN ($i.e. ~T < M_i$), the decay of the respective heavy field $N_i$ becomes relevant for generating a CP-asymmetry along a particular lepton flavor ($\alpha$) direction parametrized by, 
\begin{align}
	\varepsilon^{(i)}_{\ell_\alpha}=\frac{\Gamma(N_i\to \ell_{L_\alpha} +H)-\Gamma(N_i\to \overline{\ell}_{L_\alpha}+\overline{H})}{\sum_\alpha \Gamma(N_i\to \ell_{L_\alpha} + H)+\Gamma(N_i\to \overline{\ell}_{L_\alpha}+\overline{H})},
\end{align}
where the denominator corresponds to the total decay rate of $N_i$ at tree level, given by:
\begin{align}
	\Gamma_{N_i}&= \sum_\alpha \Gamma(N_i\to \ell_{L_\alpha} +H)+\Gamma(N_i\to \overline{\ell}_{L_\alpha} +\overline{H}) \nonumber \\
	&= \frac{(Y_{\nu}^{\dagger}Y_{\nu})_{ii}}{8 \pi} M_i.
\end{align}
The out-of-equilibrium condition necessary for lepton asymmetry production is satisfied when the decay rate of $N_i$ 
remains smaller than the expansion rate of the Universe. 

\subsection{Unflavored estimate}

A non-zero $\varepsilon^{(i)}_{\ell_\alpha}$ would follow due to the interference between the tree level and loop-level decay amplitudes. With a hierarchical RHN masses as $M_1 \ll M_2 \ll M_3$, the CP-asymmetries generated by $N_2$ and 
$N_3$ are however expected to be washed out by the lepton number violating interactions of $N_1$, leaving $\varepsilon^{(1)}_{\ell_\alpha}$ ($\equiv \varepsilon_{\ell_{\alpha}}$ where we omit the generation index $^{(1)}$ henceforth) 
as the only relevant one for leptogenesis, given by 
\begin{widetext}
\begin{align}
\varepsilon_{\ell_{\alpha}} =  \frac{1}{8 \pi (Y_{\nu}^{\dagger}Y_{\nu})_{11}} \sum_{j\neq 1}\Bigg\{\rm{Im}\left[(Y_{\nu}^*)_{\alpha 1} (Y_{\nu})_{\alpha j} (Y_{\nu}^{\dagger} Y_{\nu})_{1 j}\right] \mathbf{F}\left(\frac{M_j^2}{M_1^2}\right)
 + \rm{Im}\left[(Y_{\nu}^*)_{\alpha 1} (Y_{\nu})_{\alpha j} (Y_{\nu}^{\dagger}Y_{\nu})_{j1}\right] \mathbf{G}\left(\frac{M_j^2}{M_1^2}\right)\Bigg\}.
 \label{eq:cp}
 \end{align}
 \end{widetext}
Here $\mathbb{F}(x)= \sqrt{x}\left[1+\frac{1}{1-x}+(1+x)\ln\left(\frac{x}{1+x}\right)\right]$ and $\mathbb{G}(x)=1/(1-x)$ are the loop functions originated from both vertex and self-energy corrections to the decay of $N_1$. Conventionally the total CP asymmetry is calculated after having the flavor sum as $\varepsilon_{\ell} = \sum_{\alpha} \varepsilon_{\ell_{\alpha}}$ which results a vanishing contribution to the second term of the r.h.s of Eq.~{\eqref{eq:cp}}. However, both the terms remain important for flavored leptogenesis.

It is shown in \cite{Davidson:2002qv} that a maximum CP-asymmetry, generated from the $N_1$ decay, can be extracted from Eq.~\eqref{eq:cp} as,
\begin{align}
|\varepsilon_\ell|\lesssim \frac{3}{8\pi} \frac{M_1}{v^2}(m_3-m_1)= \varepsilon_\ell^{\text{Max}}.
\label{epsilon-limit}
\end{align}
Such an asymmetry however be washed out partially due to the lepton number violating interactions so as to write down the final $B-L$ asymmetry including the efficiency factor $\kappa_f$ by 
\begin{align}
Y_{B-L}\equiv n_{B-L}/s = -\frac{1}{7.04} \frac{3}{4} \varepsilon_\ell \kappa_f.
\label{eq:final-asymmetry}
 \end{align}
 where, $Y_x=n_x/s$ represents the number density to entropy density ratio for $x$-species.
Hence, in case the $N_1$ responsible for the asymmetry generation was in thermal equilibrium at the early Universe, the limit on CP-asymmetry as in Eq.~\eqref{epsilon-limit} leads to an estimate of maximal baryon asymmetry $Y_B^{\text{Max}}$. Requirement of $Y_B^{\text{Max}}\geq Y_B^{\rm{exp}} =  8.718 \times 10^{-11}$ \cite{Cyburt:2015mya,Planck:2018vyg} eventually provides a lower limit on lightest RHN mass \cite{Davidson:2002qv,Hahn-Woernle:2008tsk} as:
\begin{align}
 M_1&\gtrsim \frac{7.04}{0.96 \times 10^{-2}}  \frac{8\pi v^2}{3 m_3} \frac{Y_B^{\rm{exp}}}{\kappa_f}\notag\\
 &\approx \frac{6 \times 10^8~\rm{GeV}}{\kappa_f}~ \left(\frac{Y_B^{\rm{exp}}}{8.718 \times 10^{-11}}\right)~\left(\frac{0.05~\rm{eV}}{m_3}\right),
 \end{align}
 where light neutrino masses are considered to be hierarchical and consistent with neutrino oscillation data \cite{Esteban:2020cvm}. 
 
An accurate estimate for the final asymmetry or in other words the efficiency factor $\kappa_f$ would however follow if one 
solves coupled BEs that correlate the abundance of the lightest RHN with the lepton number asymmetry produced. Note that in this case, the reheating temperature (considering instantaneous reheating) $T_{\text{RH}}$ should be more than $M_1$ indicating $T_{\text{RH}} \gtrsim 10^{10}$ GeV or so. \\
 
 \subsection{Flavored Regime and Charged Lepton Equilibration}
 
In evaluating the final lepton asymmetry above, a flavor sum is performed. However, it has been found that 
the situation can actually be more complicated as soon as charged lepton Yukawa interactions ($Y_{\alpha} \overline{\ell}_{L_\alpha} H e_{R_\alpha}$ with $e_{R\alpha}$ as representative of right handed electron/muon/tau) become faster compared to $N_1-\ell_L H$ interaction \cite{Blanchet:2006ch}. In that case, during the out-of-equilibrium decay process of the $N_1$, the charged lepton Yukawa interaction for one or more flavor(s) may enter equilibrium leading to the breaking of quantum coherence of the lepton doublet state along different flavor directions produced from the $N_1$ decay \cite{Barbieri:1999ma,Nardi:2006fx,Abada:2006fw,Blanchet:2006be,Dev:2017trv}. As a result, lepton asymmetry along individual flavors may start to become distinguishable. In this case, one needs to look for the evolution of the individual 
flavor lepton asymmetries instead of total lepton number asymmetry by constructing BEs for lepton asymmetries along individual flavors. 
 
\subsubsection{Evaluation of Equilibration Temperature}

In order to check whether the charged lepton Yukawa interaction of a particular flavor $\alpha$ is fast enough at a given temperature to be in thermal equilibrium, the associated interaction rate ($\Gamma_{\alpha}$) has to be more than the expansion rate of the Universe \cite{Nardi:2006fx}. Since, this charged lepton equilibrium temperature (ET) plays a decisive role in determining the flavor effect, we elaborate on it in case of {\it thermal} leptogenesis here. In the standard scenario, assuming all these phenomena are occurring in a radiation dominated Universe, the thermally averaged interaction rates of the SM Higgs doublet decaying to left-handed lepton doublets and right-handed charged lepton singlets (more specifically $H\leftrightarrow \ell_{L_\alpha} e_{R_\alpha} $\cite{Campbell:1991at}) can be estimated as \cite{Campbell:1991at,PhysRevLett.71.2372,Cline:1993bd}:
\begin{widetext}
	\begin{align}
		\langle\Gamma_\alpha \rangle=\int \frac{d^3 p} {(2\pi)^3 2E_P}\int \frac{d^3 k}{(2\pi)^3 2E_k}\int \frac{d^3 k^\prime}{(2\pi)^3 2E_{k^\prime}}
		(2\pi)^4 \delta^{(4)}(p-k-k^{\prime})|\mathcal{M}|^2 \frac{f_p}{n_p},
		\label{eq:decayrate}
	\end{align}
\end{widetext}
where $p$ is the $4$-momentum of the Higgs while $k$ and $k^{\prime}$ are the $4$-momentum of lepton doublet and singlet right handed charged lepton respectively. The thermal distribution of Higgs $f_p$ and number density $n_p$  are taken as:
\begin{align}
	f_p =\frac{1}{e^{E_p/T}-1}\,,~
	n_p=\frac{\zeta(3) T^3}{\pi^2}\,.
\end{align}
The matrix amplitude squared $|\mathcal{M}|^2$ for such decay would be (assuming final state particles have negligible mass):
\begin{align}
	|\mathcal{M}|^2= 2 Y_{\alpha}^2 k.k^{\prime}= Y_{\alpha}^2 M_H^2\,,\quad \alpha=e,\mu,\tau.
\end{align}
Evaluation of the integrals in Eq.~\eqref{eq:decayrate} for $T \gg M_H$ yields \cite{Campbell:1991at}:
\begin{align}
	\langle\Gamma_\alpha \rangle=\frac{Y_{\alpha}^2 \pi}{192 \zeta(3)T} M_H^2.
\end{align}
Considering the thermal mass of the Higgs to be\cite{Weldon:1982bn,Quiros:1999jp,Senaha:2020mop}: 
\begin{align}
	M_H =	M_H(T)\simeq \frac{T}{4}\sqrt{3 g^2+g^{\prime^2}+ 4y_t^2+8 \lambda},
\end{align}
where $g,g^{\prime}$ are SM gauge coupling constants and $y_t$, $\lambda$ are the top Yukawa and the Higgs quartic couplings respectively, the thermally averaged interaction rate for the decay processes will become \cite{Abada:2006ea} $\mathcal{O} (5 \times 10^{-3}) Y_{\alpha}^2 T$.

However, thermal corrections for the final state particles can also be important as they are \cite{Weldon:1982bn}: $m_{\ell_L} (T)= \frac{1}{4} \sqrt{\left(3 g^2 + g'^2\right)}T$ and $m_{e_R} (T) = \frac{1}{2} g' T$ for $\ell_L$ and $e_R$ respectively. 
Though the hierarchy $M_H (T) > m_{\ell_L} (T) > m_{e_R} (T)$ is always maintained for $T > T_{EW}$, a 
situation can be achieved at some high temperature where this Higgs decay channel may actually be closed with 
$M_{H} (T)$ being smaller than $m_{\ell_L} (T) + m_{e_R} (T)$. This happens due to the decrease of top Yukawa coupling $y_t$ with the increase in temperature~\cite{Bodeker:2019ajh}. 
\begin{figure}[t]
	\includegraphics[width=1\linewidth]{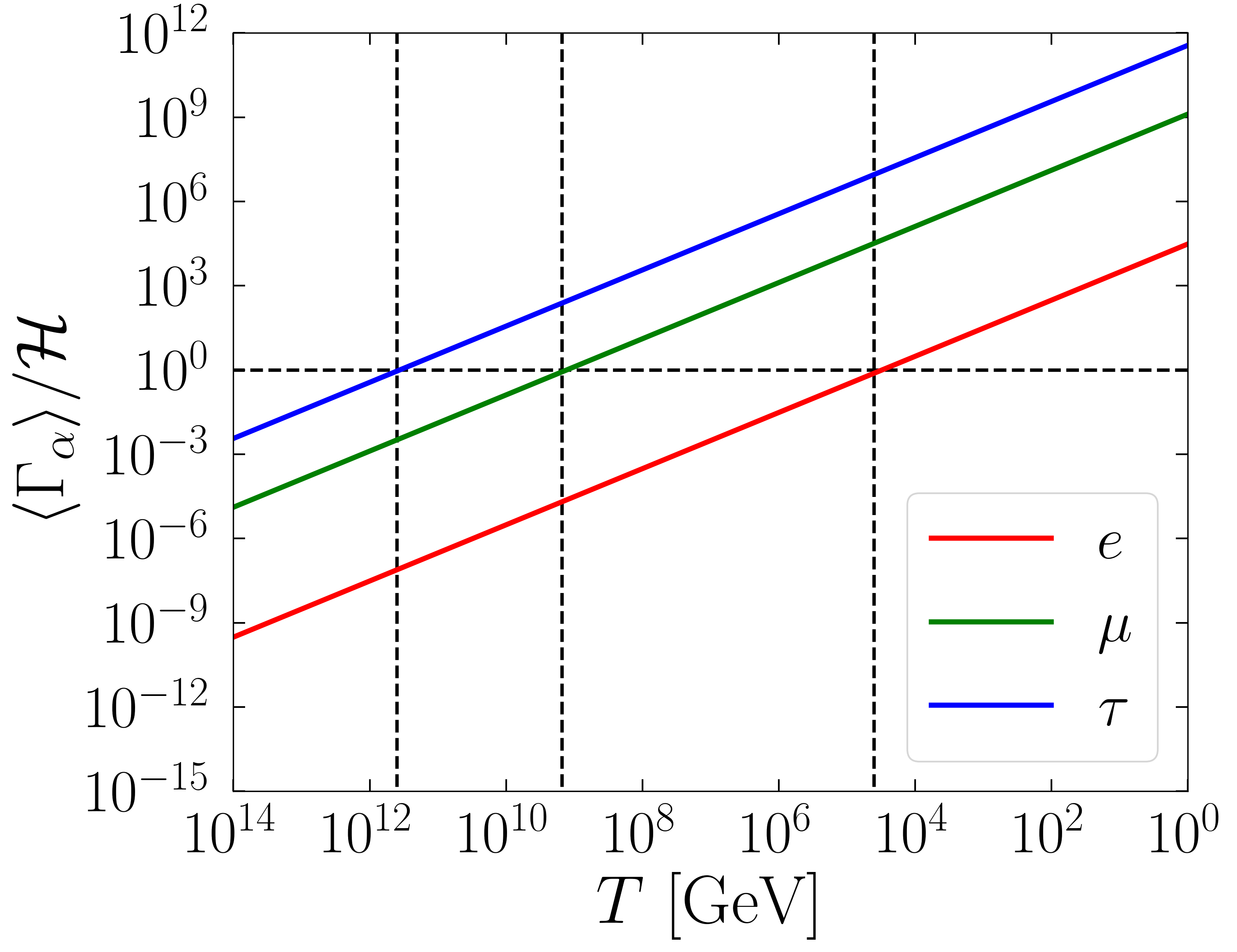}
	\captionsetup{justification=raggedright }
	\caption{Variation of $\langle\Gamma_\alpha \rangle/H$ with respect to ($w.r.t.$) $T$ in radiation dominated phase. Horizontal dashed line denotes $\langle\Gamma_\alpha \rangle/H=1$, while the vertical dashed lines indicate the ETs of three charged Yukawa interactions.}\label{fig:1}
\end{figure}
Note that apart from the 1$\rightarrow$ 2 decay, the 2$\leftrightarrow$2 scatterings involving the specific Yukawa interaction $Y_{\alpha}$ (such as $X H^{\dagger} \rightarrow \bar{\ell}_L e_R, \ell_L H^{\dagger}  \rightarrow X e_R$ $etc.$ where $X = B, W$ gauge bosons) are also important at high temperature as their contributions to the interaction rate falls in the range: $(5.19 - 4.83) \times 10^{-3} ~Y^2_{\alpha} T$ \cite{Garbrecht:2013bia,Garbrecht:2014kda} which is found to be in the same ballpark of the naive decay estimate above. Hence, for the purpose of our study, we consider the interaction rate $\langle\Gamma_\alpha \rangle$ associated to the charged lepton Yukawa $Y_{\alpha}$ to be 
\begin{align}
	\langle\Gamma_\alpha \rangle\simeq 5 \times 10^{-3} Y_{\alpha}^2 T.
	\label{eq:gamma}
\end{align}

Comparison between the obtained interaction rates $\langle\Gamma_\alpha \rangle$ with Hubble constant ($\mathcal{H}= 1.66 g_*^{1/2} T^2/M_{pl}$ in radiation dominated Universe, where $M_{pl}$ is the Planck mass.) will lead to the ET of right-handed charged lepton singlets. 
Figure \ref{fig:1} shows the variation of $\langle\Gamma_\alpha \rangle/{\mathcal{H}}$ with respect to temperature $T$ for different lepton flavors: $\alpha=e,\mu,\tau$. Note that $\tau$ Yukawa interaction becomes fast enough around $T = T^*_{0(\tau)}\simeq 5 \times 10^{11}$ GeV (evident from the intersection of $\langle\Gamma_\tau \rangle/{\mathcal{H}}$ line in blue with 1) while muon Yukawa interaction comes to equilibrium at $T^*_{0(\mu)} \simeq 10^9$ GeV as seen from $\langle\Gamma_\mu \rangle/{\mathcal{H}} =1$ point of Fig.~\ref{fig:1}.

\subsubsection{Effects of Flavor and Boltzmann Equations}

As a result, if a $N_1$ decays while staying in out-of-equilibrium in a temperature range $10^{9}\lesssim T\lesssim  5 \times 10^{11}$ GeV, lepton asymmetry becomes distinguishable along two orthogonal directions denoted by $\alpha = a$ (specifying a coherent superposition of $e$ and $\mu$ lepton flavors) and $\tau$. Hence contrary to the unflavored case, here we need to study the evolution of $B/3-L_{\alpha} \equiv \Delta_{\alpha}$ charges with $\alpha = a$ and $\tau$. With a further reduction of the temperature below $T\lesssim 10^9$ GeV, the lepton doublets completely loose their quantum coherence. Hence, at this stage, lepton asymmetry becomes distinguishable along all three flavor directions $e$, $\mu$  and $\tau$. As 
a result, evolution of lepton asymmetry along all three directions becomes relevant to produce final baryon asymmetry.

Incorporating the flavor effects into account, the evolution of the abundance of decaying $N_1$s ($Y_{N_1}$) and lepton asymmetries along individual flavor direction ($Y_{\Delta_{\alpha = e, \mu, \tau}}$) can be represented by the following sets of flavored classical BEs (neglecting $\Delta L=1,2$ scattering processes and subtracting the on-shell contribution from $\Delta L=2$ process) \cite{Nardi:2005hs,Nardi:2006fx}:
\begin{widetext}
	\begin{align}
		s H z \frac{d Y_{N_1}}{dz}
		&=
		-
		\left( \frac{Y_{N_1}}{Y_{N_1}^{\rm eq}}	-	1	\right)
		\gamma_{D}  \,,\label{be1}
		\\
		s H z \frac{d Y_{\Delta_{\alpha}}}{dz}
		&=
		-\Bigg\{
		\left(	\frac{Y_{N_1}}{Y_{N_1}^{\rm eq}}	-1\right)
		\varepsilon_{\ell \alpha} \gamma_{D}
		+
		K^0_{\alpha} \sum_{\beta} \Bigg[ \frac{1}{2}  (C^{\ell}_{\alpha \beta} +
		C^H_{\beta}) \gamma_D \Bigg]\frac{Y_{\Delta_{\beta}}}{Y_{\ell}^{\rm eq}}
		\Bigg\},\label{eq:lep}
	\end{align}
\end{widetext}
where $K^0_{\alpha}= \frac{(Y_{\nu}^*)_{\alpha 1} (Y_{\nu})_{\alpha 1}}{(Y_{\nu}^{\dagger} Y_{\nu})_{11}}$ is known as flavor projector \cite{Nardi:2006fx,Blanchet:2006be} and $C^{\ell}, C^H$ matrices connect the asymmetries in lepton and Higgs sectors to asymmetries in $\Delta_{\alpha}$ expressed in terms of $Y_{\Delta{\alpha = e, \mu, \tau}}$ or $Y_{\Delta{\alpha = a, \tau}}$ depending on the leptogenesis scale \cite{Nardi:2006fx}. Eq. \eqref{eq:lep} can be a set of two (three) equations if $10^9<M_1<5 \times 10^{11}$ GeV ($M_1< 10^9$ GeV). Here $\gamma_D$ represents the total decay rate density of $N_1$: 
\begin{align}
	\gamma_D= n_{N_1}^{\rm{eq}}\langle \Gamma_{N_1}\rangle, \qquad \langle\Gamma_{N_1} \rangle= \frac{K_1(z)}{K_2(z)}\frac{(Y_{\nu}^{\dagger} Y_{\nu})_{11}}{8\pi} M_1\,,
	\label{eq:17}
\end{align}
where $K_1(z)$ and $K_2(z)$ are the modified Bessels functions. $z=M_1/T$ is a dimensionless quantity, with respect to which we will look for the evolution of $Y_x$. The equilibrium number density of the $N_1$ can be expressed as:
\begin{align}
	n_{N_1}^{\rm{eq}}=\frac{g M_1^3}{2 \pi^2 z} K_2\left(z\right),
\end{align}
where $g$ is the number of degrees of freedom of $N_1$.

To study the evolutions numerically, we need to provide inputs for the neutrino Yukawa coupling $Y_{\nu}$ matrix the structure of which can be extracted using Casas-Ibarra (CI) parametrization \cite{Casas:2001sr} as:
\begin{align}
	Y_{\nu}=-i \frac{\sqrt{2}}{v} U D_{\sqrt{m}} \mathbf{R} D_{\sqrt{M}}\,,
	\label{CI}
\end{align}
where $U$ is the Pontecorvo-Maki-Nakagawa-Sakat (PMNS) matrix \cite{Zyla:2020zbs} which connects the flavor basis with mass basis for light neutrinos. $D_{\sqrt{m}}=  \rm{diag}(\sqrt{m_1},\sqrt{m_2},\sqrt{m_3}) $ is the diagonal matrix containing the square root of light neutrino mass and similarly $D_{\sqrt{M}}= \rm{diag}(\sqrt{M_1},\sqrt{M_2},\sqrt{M_3})$ represents the diagonal matrix for RHN masses. $\mathbf{R}$ is an orthogonal matrix satisfying $\mathbf{R}^{\rm{T}}\mathbf{R}=1$. 

\begin{figure}[h]
	\includegraphics[width=1\linewidth]{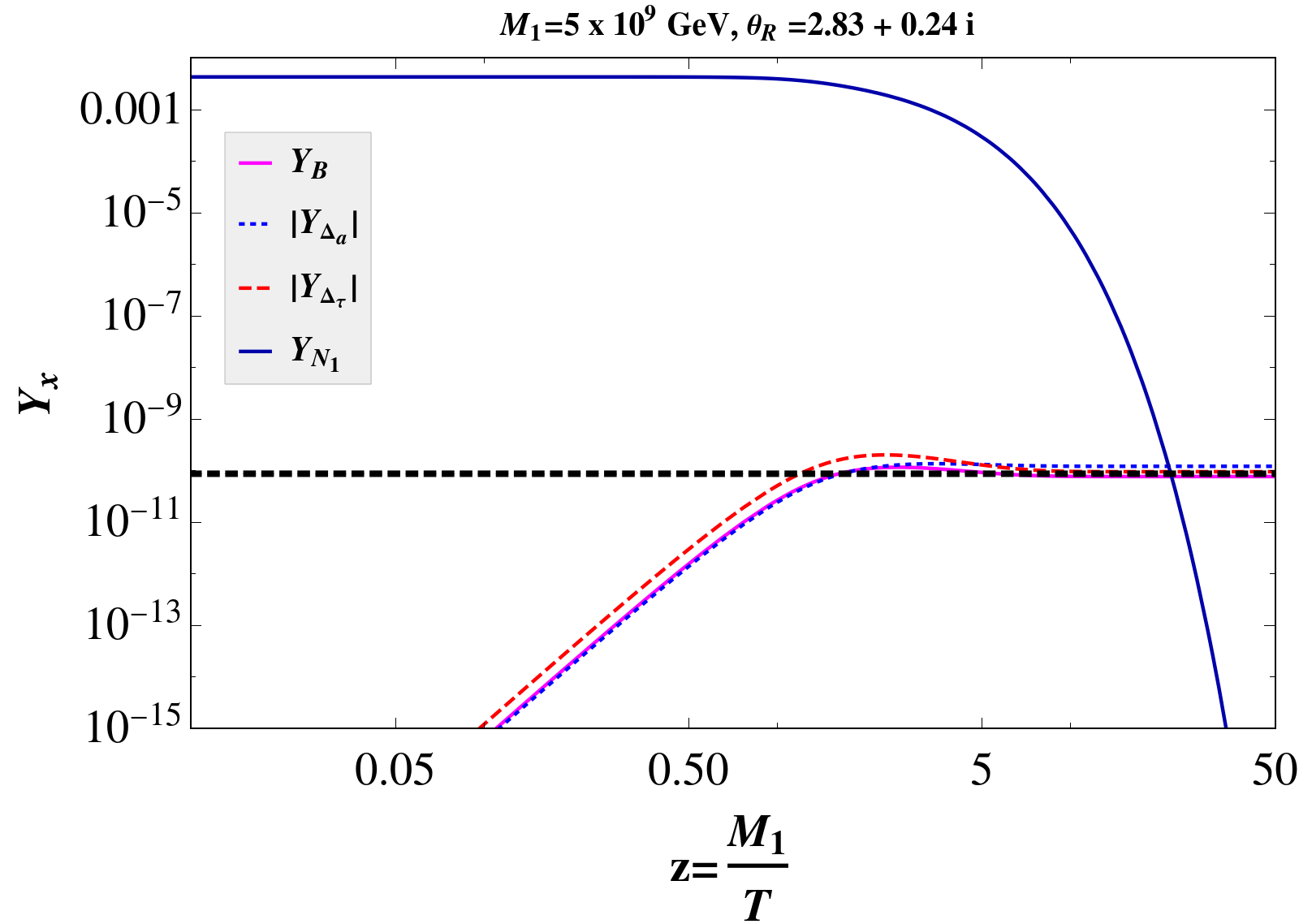}
	\captionsetup{justification=raggedright }
	\caption{ Evolution of various comoving number densities with respect to $z$ for $M_1= 5\times 10^9$ GeV. Horizontal black dashed line indicates the observerd baryon asymmetry $Y_B^{\text{exp}}$.}\label{fig:2}
\end{figure}

Fig.~\ref{fig:2} depicts a scenario where evolution of $ \tau$ and $a$ lepton flavor become essential as the mass of the lightest RHN is chosen to be $M_1=5 \times 10^9$ GeV. While constructing the $Y_\nu$ for this case, we consider a typical hierarchy among the RHNs as $M_3 = 100 M_2, ~M_2=100M_1$ along with we use the best fit values of solar and atmospheric mass-squared differences (considering $m_1 =0$) and mixing angles, and CP-violating phase $\delta$ to define $U$ via Eq.~\eqref{CI}. For such a set of hierarchical RHNs, $\mathbf{R}$ can be considered 
to have the following structure:
\begin{align}
	\mathbf{R}=\begin{pmatrix}
		0 & \cos \theta_R & \sin \theta_R\\
		0 & -\sin \theta_R & \cos \theta_R \\
		1 & 0  & 0
	\end{pmatrix},
\end{align}
Where $\theta_R$ can be, in general, a complex angle, chosen to be  $\theta_R=2.83+0.24 i$ for this case so as to realize the final asymmetry to be consistent with the correct baryon asymmetry of the Universe. As can be seen from the Fig.~\ref{fig:2}, starting from a symmetric Universe, lepton asymmetries along $a$ and $\tau$ direction (blue dot and red dashed lines) start to grow due to out-of-equilibrium decay of lightest RHN (the corresponding abundance of $N_1$ is indicated by solid blue line) and saturates around $z\sim 2$. Similar behavior can be seen for the number density of baryon asymmetry (magenta line) as well. Because of the sphaleron processes, produced lepton asymmetries get converted to baryon asymmetry via the relation: $Y_B= 28/79 \sum_{\alpha} Y_{\Delta_{\alpha}}$. Eventually with large value of $z$, this baryon asymmetry saturates to experimentally observed baryon asymmetry: $Y_{B}^{\rm{exp}} = (8.718 \pm 0.012)\times 10^{-11}$ \cite{Cyburt:2015mya,Planck:2018vyg} (dictated by black dashed lines of Fig.~\ref{fig:2}). The discussion in this section embarks on the importance of flavor aspects in {\it thermal} leptogenesis which is now being explored in the context of non-instantaneous reheating period in this work. Before that, we discuss the {\it non-thermal} leptogenesis in brief. 

\section{RHNs Produced from Inflaton Decay}
\label{section:3}

We may now turn our attention to a situation where the RHNs are being produced from inflaton ($\phi$) decay. In case the inflaton decays solely to $N_1$s having decay width $\Gamma_\phi$, the radiation component of the Universe arises as a result of further decay of those RHNs (with decay width $\Gamma_{N_1}$) into SM lepton and Higgs doublets which thermalize rapidly. With $\Gamma_\phi << \Gamma_{N_1}$, the reheating temperature assuming instantaneous reheating\footnote{Here contribution from preheating \cite{Kofman:1994rk,Kofman:1997yn,Greene:1998nh} is not taken into account.} is governed by the condition: $\Gamma_\phi= \mathcal{H}$ given by:
  \begin{align}
 T_{\text{RH}}=\left(\frac{45}{4 \pi^3 g_*} \right)^{1/4}\sqrt{\Gamma_\phi M_{pl}}\,.
 \end{align}
where $g_*$ represents the effective number of relativistic degrees of freedom in the SM.
In this case, as $T_{\text{RH}} < M_1$, the $N_1$s would decay immediately after being produced from inflaton. 

The decay of such non-thermally produced $N_1$ also generates a lepton asymmetry which can be converted to baryon asymmetry (provided $T_{\text{RH}}>100$ GeV), expressed as~\cite{Lazarides:1991wu,Asaka:1999yd, HahnWoernle:2008pq}
 \begin{align}
 Y_B=-\frac{28}{79}\frac{n_{B-L}}{s}= -\frac{28}{79} \varepsilon_{\ell} \frac{n_{N_1}}{s}=-\frac{28}{79} \frac{3}{4}\varepsilon_{\ell}\frac{T_{\text{RH}}}{m_\phi}\,. 
 \end{align}
Here we assume $\phi$ decays to $N_1$ only. Also, washout by the lepton number violating process turns out to be insignificant. The additional suppression factor, proportional to $T_{\text{RH}}/m_{\phi}$ is related to the ratio of the number 
density of produced $N_1$ and entropy $s$ and follows from the equality $\rho_{\phi} = \rho_R$ at $T=T_{\text{RH}}$. 
Note that, due to the condition $M_1> T_{\text{RH}}$, the inverse decay process $\ell + H \rightarrow N_1$ can not take place and consequently, the loss of flavor coherence is not expected to occur. However, the situation alters once we consider an extended period of reheating as we plan to discuss next. 

\section{flavor effect during reheating period}
\label{section:4}
As previously discussed, a {\it thermal} leptogenesis scenario with $N_1$ mass $M_1\lesssim  5 \times 10^{11}$ GeV experiences flavor effects since the charged lepton Yukawa interactions start to enter equilibrium below this temperature. An estimate of such ETs for different flavors is provided in Fig.~\ref{fig:1}. Note that the ETs of these Yukawa interactions (associated to different flavors of right handed charged leptons) are calculated assuming that the Universe is already in a radiation domination era followed by an instantaneous reheating after inflation. Obviously, such consideration leads to $T_{\text{RH}} > M_1$ indicating the presence of high reheating temperature. Now, it is also known that the reheating might not be an instantaneous process \cite{Chung:1998rq,Giudice:2000ex,Garcia:2020eof,Haque:2020zco}. On top of that, the reheating temperature can be low enough (though larger than few MeV from BBN limit \cite{Giudice:2000ex,Kawasaki:2000en,Martin:2010kz,Dai:2014jja}). 

The era of reheating begins when after inflation, the inflaton field $\phi$ starts to decay. Neglecting the possibility via preheating \cite{Kofman:1994rk,Kofman:1997yn,Greene:1998nh}, as this field $\phi$ starts to decay to the lighter SM degrees of freedoms, the thermalization of these light decay products helps the Universe to attain a maximum temperature $T_{\text{Max}}$. Subsequently, the temperature of the Universe falls at a rate much slower than the standard scaling $T\sim a^{-1}$ ($a$ is the Friedmann–Robertson–Walker scale factor). This continues till a point (defining $T_{\text{RH}}$) where the radiation component becomes the dominating one in the Universe. This nontrivial behavior of the temperature as function of scale factor $a$
results due to the faster expansion rate compared to the standard scenario (quantified by Hubble $\mathcal{H}$) during the lifetime of the inflaton. 

As a result of this modified $\mathcal{H}$, a change in the behaviour of $\langle \Gamma_\alpha \rangle/{\mathcal{H}}$ is also expected in the period in between $T_{\text{Max}}$ and $T_{\text{RH}}$, compared to the standard scenario represented in section \ref{section:2} as estimated in \cite{Datta:2022jic} by the present authors. This prolonged reheating was dictated by the size of the effective coupling of the inflaton with SM fields. This would be particularly prominent provided $T_{\text{RH}}$ falls below $T^*_{0(\tau)}$ and $T_{\text{Max}}$ maintains a relation $T_{\text{Max}} > T^*_{0(e)}$. In fact, as a result of such a delayed entry (due to the modified $\mathcal{H}$ for $T_{\text{Max}} > T > T_{\text{RH}}$) of the charged lepton Yukawa interactions into the equilibrium, a shift of the flavor regime(s) of leptogenesis compared to the standard scenario is expected in this case of extended reheating period. 

If the lepton asymmetry is generated from the out-of-equilibrium decay of the lightest RHN, then depending on the scale of leptogenesis, three possibilities can be realized in presence of this extended reheating: (A) $M_1 < T_{\text{RH}}$, (B) $T_{\text{Max}} > M_1 > T_{\text{RH}}$ and (C) $M_1 > T_{\text{Max}}$. Case (A) corresponds to the standard {\it thermal} leptogenesis as by the time decay of $M_1$ becomes effective for leptogenesis, Universe is already in radiation dominated era. So, effect of this extended period of reheating does not carry any additional impact here. On the other hand, with case (C), $N_1$ can never be produced thermally. They can only be produced non-thermally from the decay of some heavy particle (provided that coupling exists), like inflaton resembling the case of purely {\it non-thermal} leptogenesis as discussed in section \ref{section:3} where the effect of flavor does not play any vital role. 

The case (B) is however the most interesting one. In \cite{Datta:2022jic}, it was shown that the lightest RHN can be produced from the thermal bath during $T_{\text{Max}} > T \gtrsim M_1$ while it decays thereafter ($T < M_1$) into SM lepton and Higgs doublets, hence generating the lepton asymmetry. Now, during this extended period of reheating, the shift in the ET of right handed charged leptons affects the lepton asymmetry production. For example, we have shown \cite{Datta:2022jic} that with an effective coupling of inflaton to SM fermions ($y_{\phi ff}$)
 of order $10^{-4}$, the ET of $\tau_R$ reduces by almost an order 
of magnitude compared to the standard case. Such inflaton-SM fermion coupling also sets\footnote{Note that there exists another parameter $\lambda$, involved in the inflation potential, which also takes part in determining these $T_{\text{Max}}$ and $T_{\text{RH}}$ values. However $\lambda$ is fixed so as to have correct prediction for the normalization of CMB anisotropies. The present analysis is also based on the same inflationary potential considered in \cite{Datta:2022jic} as we elaborate upon soon.} $T_{\text{Max}}  \sim 7 \times 10^{11}$ GeV while $T_{\text{RH}}$ becomes $4 \times 10^{10}$ GeV. 
Consequently,  $N_1$ of mass $M_1 \simeq 10^{11}$ GeV can be produced thermally during reheating (as $T_{\text{Max}} \sim 7 \times 10^{11}$ GeV) while it decays around $T \lesssim M_1$. The shift in the ET renders the corresponding leptogenesis as an unflavored case which otherwise falls in the ballpark of flavored leptogenesis (two flavor regime). 

Motivated by the above result, in this work, we further consider an intriguing extension (in the context of case (B) itself) where in addition to the inflaton-SM fermion effective coupling ($y_{\phi ff}$) there exists a direct coupling ($y_{\phi N N}$) between the inflaton and $N_1$s. Introduction of such a coupling not only induces $N_1$s in the system from the decay of the inflaton (in addition to the thermally generated $N_1$), but also opens up the possibility of modifying the Hubble further, hence affecting $T_{\text{RH}}$ as well as the ET depending on its relative coupling strength compared to inflaton-SM fermion coupling. 

Hence, with a nonzero branching of the inflaton to $N_1$ in addition to the inflaton-SM fermion one, we expect to have $N_1$ production from inflaton decay as well as thermal production of it via inverse decay for a temperature range $T_{\text{Max}} > T \gtrsim M_1$. These $N_1$ however find themselves in out of equilibrium as the Hubble $\mathcal{H}$ at this temperature regime remains large enough ($\phi$ dominates). Therefore, $N_1$ would decay and may contribute 
to lepton asymmetry generation even at temperature above $M_1$ unless it has been washed out by inverse decay. In case $\rho_{N_1}$ dominates over $\rho_{R}$, the washout by inverse decay turns out to be weak so as not to erase the asymmetry. This particular era of leptogenesis turns out to be somewhat different from a purely thermal or non-thermal one 
since the lepton asymmetry here is generated from the decay of both the thermally produced $N_1$s and those produced from inflaton decays in this regime. We call it as `{\it quasi}'-thermal leptogenesis.
Additionally for temperature below $T \sim M_1$, leptogenesis proceeds in the usual way. However, with a significantly dominant coupling of $y_{\phi NN}$ over $y_{\phi ff}$, $N_1$s can even be produced beyond $T = M_1$ point. In this case ($T < M_1$), such non-thermally produced $N_1$ would instantaneously decay and contribute to lepton asymmetry production similar to the usual {\it non-thermal} leptogenesis scenario.   

Following the above discussion, we now construct the relevant Lagrangian (apart from the SM one) as given by,
\begin{align}
	-\mathcal{L} =  y_{\phi ff}\phi \overline{f} f+ y_{\phi NN}\phi \overline{N^c_1} N_1 + V(\phi),
	\label{eq:l1}
\end{align}
in addition to the Type-I seesaw Lagrangian of Eq.~\eqref{eq:1}. Here, $y_{\phi ff}$ is only an effective coupling and $f(\overline{f})$ are the SM fermions. For simplicity, we only keep the coupling of inflaton with 
the lightest $N_1$. Here, $V(\phi)$ corresponds to the scalar potential of inflaton $\phi$ describing 
the post-inflationary epoch during which the inflaton field oscillates about the minimum (origin), relevant for realizing reheating. We have taken a power-law form for $V(\phi)$ about the minima \cite{Garcia:2020eof} as:
\begin{align}\label{eq:Vgen}
	V(\phi) = \lambda \frac{\left|\phi\right|^n}{M_P^{n-4}}\, ,
\end{align}
where $M_P$ is the reduced Planck mass. The magnitude of the coupling $\lambda$ can be estimated from the CMB observables such as spectral index and tensor-to-scalar ratio and depends on the order of the polynomial $n$. Origin of such choice of potential near minimum ($\phi = 0$) can be traced back to T-attractor models of inflation in no-scale supergravity \cite{Kallosh:2013hoa, Khalil:2018iip}, 
\begin{align}
		V_{\rm{Inf}}(\phi)= \lambda M_P^4 \left[\sqrt{6}\tanh\left(\frac{\phi}{\sqrt{6}M_P}\right)\right]^n,
		\label{eq:vphi}
	\end{align}
such that the $V(\phi)$ of Eq. (\ref{eq:Vgen}) follows after expanding the above $V_{\rm{Inf}}(\phi)$ about the origin ($\phi \ll M_P$) to the leading order.


\vskip 0.3cm
After the end of the inflation, the inflaton starts to perform damped oscillations about the minima of the potential and eventually decays to the SM fermion-antifermion pair as well as to $N_1$ following Eq. ~\eqref{eq:l1}. Here we ignore any potential contribution that may come from preheating \cite{Greene:1998nh,Giudice:1999fb}. Consequently, 
the energy density of the inflaton field $\rho_{\phi}$ satisfies the equation:
\begin{align}
	\frac{d \rho_\phi}{dt} 
	+ 3 \left(\frac{2n}{n+2}\right)\mathcal{H} \rho_\phi &= - (\Gamma_{\phi ff}+\Gamma_{\phi NN}) \rho_\phi.
	\label{Eq:eqrhophi}
\end{align}
Here the term proportional to $\mathcal{H}$ indicates the dilution of energy density due to the expansion of the Universe while the term on the right hand side of the BE represents the dilution (hence comes with a negative sign) of the energy density of the $\phi$ as a result of its decay to $N_1$ and SM fermion/anti-fermions. In the right hand side,
$\Gamma_{\phi ff}$  and $\Gamma_{\phi NN}$ represent the decay widths of inflaton to SM fermions and $N_1$ respectively and are expressed as 
\begin{align}
	\Gamma_{\phi ff} &= \frac{y_{\phi ff}^2}{8 \pi} m_\phi\,,\quad \Gamma_{\phi NN} = \frac{y_{\phi NN}^2}{8 \pi} m_\phi\, .
	\label{Eq:eqgammaphi}
\end{align}
 At this place, it is pertinent to discuss on the inflaton mass $m_{\phi}$. In such setups, the effective inflaton mass $m_\phi$ (= $[\partial^2_\phi V(\phi)]^{1/2}$ at minimum) 
 of the oscillating inflaton field $\phi(t)$ becomes a function of the inflaton field itself, since
\begin{align}
\partial^2_\phi V(\phi)=\; \lambda ~n(n-1) \phi^{n-2} M_P^{4-n}.
	\label{Eq:mphi} 
	\end{align}
For $n=2$, a definite mass is associated with the inflaton and a perturbative decay of the inflaton happens naturally via the $y_{\phi ff}$ and $y_{\phi NN}$ interactions. We restrict ourselves with $n=2$ in this work. However, for $n > 2$, though the inflaton is apparently massless at origin, a condensate of the oscillating inflaton seems plausible \cite{Garcia:2020eof, Garcia:2020wiy} for which a time-dependent effective mass of the oscillating inflaton can be associated as, 
\begin{align}
		m_\phi^2 = n(n-1)M_P^{\frac{2(4-n)}{n}}\lambda^{\frac{2}{n}}\rho_\phi^{\frac{n-2}{n}},
\end{align} 
where $\rho_{\phi}$ is the energy density of the inflaton field $\phi$. This is obtained after taking average of the equation of motion of the $\phi(t)$ field over one complete oscillation under envelope approximation\footnote{ The envelope approximation can be stated as $\phi(t) = \phi_0 (t) \mathcal{A}(t)$. Here $\phi_0(t)$ describes the envelope-function representative of red-shift and decay in a longer time scale while $\mathcal{A}$ stands for oscillatory behaviour (within the envelope) of it at short time scale such that $\rho_{\phi} = V(\phi_0)$ (for more details, see \cite{Shtanov:1994ce,Garcia:2020eof,Garcia:2020wiy}).}, as suggested in \cite{Shtanov:1994ce,Garcia:2020eof}.

The produced fermion-antifermion pairs would interact quickly among themselves to produce other SM particles and rapidly thermalizes producing the radiation energy density component $\rho_R$. At this stage, we can define the temperature of the Universe via 
\begin{align}
	T= \left[\frac{30 \rho_R}{\pi^2 g_*}\right]^{1/4}.
	\label{eq:temp}
\end{align}
On the other hand, the  $N_1$s produced from the inflaton decay further decays to the SM particles which will eventually contribute to $\rho_R$ too. Additionally, as per our consideration in case (B), the thermal bath can also produce back $N_1$ particularly for the temperature of the Universe $T_{\text{Max}} > T  \gtrsim M_1$. The BEs for $\rho_{N_1}$ and $\rho_R$ can therefore be written as, 
\begin{align}
	\frac{d \rho_{N_1}}{dt} 
	+ 3\mathcal{H} \rho_{N_1} &=-(\rho_{N_1}-\rho_{N_1}^{eq})\langle \Gamma_{N_1}\rangle + \Gamma_{\phi NN} \rho_{\phi}\,,
	\label{Eq:eqrhoN}\\
	\frac{d \rho_R}{dt} 
	+ 4\mathcal{H} \rho_R&=(\rho_{N_1}-\rho_{N_1}^{eq}) \langle\Gamma_{N_1}\rangle + \Gamma_{\phi ff} \rho_\phi.
	\label{Eq:eqrhor}
\end{align}
In all the BEs above, $\mathcal{H}$ represents the Hubble expansion rate to be written as 
\begin{align}
	\mathcal{H}^2=\frac{ \rho_{\phi}+\rho _{N_1}+\rho_{R}}{3 M_{P}^2},
	\label{eq:hubble}
\end{align} 
since in this epoch $T_{\text{Max}} > T > T_{\text{RH}}$, the energy density of the Universe comprises of the components $\rho_{\phi}, \rho_{N_1}$ and $\rho_R$.  
The $\rho_{N_1}^{\rm{eq}}$ is the equilibrium energy density of $N_1$ as given by
\begin{align}
	\rho_{N_1}^{\rm{eq}}= \frac{M_1^4\left[\frac{3}{z^2} K_2(z)+\frac{1}{z} K_1(z)\right]}{\pi^2},
\end{align}
where $K_1, K_2,$ and $z$ have already been defined  while explaining Eq.~\eqref{eq:17}. The presence of this term in the above BEs is related to the existence of the inverse decay from radiation bath to produce $N_1$. Eq.\eqref{Eq:eqrhophi}, \eqref{Eq:eqrhoN}-\eqref{Eq:eqrhor} therefore together represent the most general set of equations to study the scenario under consideration. 

After discussing the $N_1$ production and the related BEs to study the respective components of the energy densities of the Universe, we now turn our attention to construct the BEs relevant for leptogenesis. As discussed earlier, being Majorana particle, the decay of the lightest RHN $N_1$ is a lepton number violating one and can produce CP asymmetry, which will eventually generate lepton asymmetry of the Universe. In order to take care effects of charged lepton Yukawa equilibration of different flavors, the following classical flavored BE can be constructed:
\begin{align}
	\frac{d n_{\Delta_\alpha}}{dt} +3 \mathcal{H} n_{\Delta_\alpha}\;=\;-\langle\Gamma_{N_1}&\rangle \Big[\frac{\varepsilon_{\ell_\alpha}}{M_1}(\rho_{N_1}-\rho_{N_1}^{\rm{eq}}) +\frac{1}{2} K^0_\alpha \notag\\
	&\times \sum_{\beta}(C^{\ell}_{\alpha \beta}+C^{H}_\beta)\frac{n_{N_1}^{\rm{eq}}}{n_{\ell}^{\rm{eq}}}n_{\Delta_\beta}\Big].
	\label{eq:bau2}
\end{align}
The equation remains identical to the one of Eq.~\eqref{eq:lep} except the fact that 
$\mathcal{H}$ is now comprised of all the contributions from inflaton, $N_1$, and radiation energy densities in line with Eq. \eqref{eq:hubble}. This will certainly influence the lepton asymmetry production differently compared to the scenario discussed 
in Section~\ref{section:2}. The first term (within the first parenthesis) on the right hand side of Eq.~\eqref{eq:bau2} represents the production of lepton asymmetry from the decay of lightest RHN $N_1$ while the remaining terms denote the washout of the produced asymmetry along individual lepton directions due to the inverse decay of the $N_1$.
Apart from the flavored setup, a situation may arise where flavor effects are not that important in discussing leptogenesis. In that case, an unflavored scenario exhibits where the evolution of the $B-L$ asymmetry is governed by the single BE given by: 
\begin{align}
	\frac{d n_{B-L}}{dt}+3 \mathcal{H}n_{B-L}=
	-\langle\Gamma_{N_1}\rangle \left[\frac{\varepsilon_\ell}{M_1}(\rho_{N_1}-\rho_{N_1}^{\rm{eq}})
	+\frac{n_{N_1}^{\rm{eq}}}{2 n_\ell^{\rm{eq}}}n_{B-L}\right].
	\label{eq:bau1}
\end{align}

Solving Eq.~\eqref{Eq:eqrhophi},~\eqref{Eq:eqrhoN},~\eqref{Eq:eqrhor} and ~\eqref{eq:bau2} or ~\eqref{eq:bau1} simultaneously, will lead to the evolution of energy density of relevant elements of the Universe and produced lepton asymmetry from the time of the end of inflaton till today. However, while solving these BEs, it is convenient to use new variables \cite{Giudice:2000ex,HahnWoernle:2008pq} for which we use the following transformations:
\begin{align}
	E_{\phi}&= \rho_\phi a^3\,,\quad E_{N_1}=\rho_{N_1} a^3\,,\quad R= \rho_R a^4\,,\notag\\ 
	&N_{B-L}= n_{B-L}a^3\,, \quad    N_{\Delta_\alpha}= n_{\Delta_\alpha}a^3
	\label{eq:newv}
\end{align}
Moreover, it is convenient to write the BEs as functions of the scale factor $(a)$ rather than time ($t$). More
precisely, we use the ratio of the scale factor to its value at the end of inflation,
\begin{align}
	A=\frac{a}{a_{\rm{end}}}.
	\label{eq:newscale}
\end{align}
Using the newly introduced dimensionless variables, BEs in Eq.~\eqref{Eq:eqrhophi},~\eqref{Eq:eqrhoN},~\eqref{Eq:eqrhor}, ~\eqref{eq:bau2} and ~\eqref{eq:bau1} will look like:
\begin{align}
	&\frac{dE_{\phi}}{dA}= 3 \left(\frac{2-n}{n+2}\right)\frac{E_\phi}{A}- \frac{(\Gamma_{\phi ff}+\Gamma_{\phi NN}) E_{\phi}}{A \mathcal{H}}\,,\label{Eq:ephi}\\
	&\frac{dR}{dA}= \frac{\langle\Gamma_{N_1}\rangle a_{\rm{end}}}{\mathcal{H}} (E_{N_1}-E_{N_1}^{\rm{eq}})+ \frac{\Gamma_{\phi ff} E_\phi}{\mathcal{H}}\,,\label{Eq:r}\\
	&\frac{dE_{N_1}}{dA}=\frac{\Gamma_{\phi NN} E_\phi}{A\mathcal{H}}-\frac{\langle\Gamma_{N_1}\rangle}{A \mathcal{H}} (E_{N_1}-E_{N_1}^{\rm{eq}}) \,,\label{Eq:en}\\
	&\frac{dN_{\Delta_\alpha}}{dA}= -\frac{\langle\Gamma_{N_1}\rangle}{A \mathcal{H}}\Bigg[\frac{\varepsilon_{\ell_\alpha}}{M_1}(E_{N_1}-E_{N_1}^{\rm{eq}})+\frac{1}{2}K^0_\alpha  \notag\\
	&\hspace{3cm}\times \sum_{\beta}(C^{\ell}_{\alpha\beta} + C^{H}_\beta) \frac{Y_{N_1}^{\rm{eq}}}{Y_\ell^{\rm{eq}}}N_{\Delta_\beta}\Bigg]
	\label{eq:bau3}\,,\\
	&\frac{dN_{B-L}}{dA}= -\frac{\langle\Gamma_{N_1}\rangle}{A\mathcal{H}} 
	\left[\frac{\varepsilon_{\ell}}{M_1}(E_{N_1}-E_{N_1}^{\rm{eq}})+
	\frac{Y_{N_1}^{\rm{eq}}}{2Y_\ell^{\rm{eq}}}N_{B-L}\right].
	\label{eq:bau4}
\end{align}
Finally, the produced lepton asymmetry can be converted to baryon asymmetry using the relation:
\begin{align}
	Y_B= \frac{28}{79} \frac{1}{s A^3} N_{B-L} = \frac{28}{79} \frac{1}{s A^3} \sum_\alpha  N_{\Delta_\alpha}.
	\label{eq:ltob}
\end{align}

\section{Results}
\label{section:5}

\begin{table}[b!]
	\begin{tabular}{||c|c|c|c|c||} 
		\hline
		Case & $y_{\phi NN}$  &  $T_{\text{RH}}$ (GeV)&  $T^*_{\tau}$ (GeV) & $Y_B$\\ [0.5ex] 
		\hline\hline
		I& $0$   & $1.67 \times 10^{10}$  &  $4.7 \times 10^{10}$ &    $8.72 \times 10^{-11}$\\
		\hline
		II &$10^{-7}$  & $1.67 \times 10^{10}$  &  $4.7 \times 10^{10}$ &   $8.72 \times 10^{-11}$\\
		\hline
		III &$10^{-5}$  & $1.72 \times 10^{10}$  &  $4.8 \times 10^{10}$ & $3.52 \times 10^{-10}$\\
		\hline
		IV & $4 \times 10^{-5}$ & $2.37\times 10^{10}$  &  $5 \times 10^{10}$ &  $4.29 \times 10^{-10}$\\
		\hline
	\end{tabular}
	\captionsetup{justification=raggedright }
	\caption{Values of $T_{\text{RH}},~ T^*_{\tau}$ and final BAU $Y_B$ for different choices of $y_{\Phi NN}$. Here we have fixed $y_{\phi ff}=4 \times 10^{-5}$ and $M_1= 6 \times 10^{10}$ GeV. $T_{\text{Max}}= 4.45 \times 10^{12}$ GeV is found to be constant for all four cases.\label{tab:table1}}
\end{table}

We now employ the set of BEs Eq.~\eqref{Eq:ephi}-\eqref{Eq:en} simultaneously in order to estimate the 
individual components of energy densities such as $\rho_{\phi}, \rho_R$ and $\rho_{N_1}$ which are connected to $E_{\phi}, E_R$ and $E_{N_1}$ respectively via Eq. \eqref{eq:newv}. By knowing $\rho_R$ as a function of the scale factor $a$ or the rescaled one $A$, the temperature can be defined by Eq. \eqref{eq:temp}. Then, using Eq. \eqref{eq:hubble}, we estimate the shift of the ET if any from their standard estimate (see Fig.~\ref{fig:1}) by comparing the interaction rate of charged lepton Yukawa interaction of individual flavor $\langle \Gamma_{\alpha} \rangle$ with $\mathcal{H}$. Afterward, depending on the shift of ET of individual flavor, we proceed for evaluating the flavored (unflavored) $B-L$ asymmetries by solving Eq. \eqref{eq:bau3} (Eq. \eqref{eq:bau4}) where we also feed the solutions of other Eqs. ~\eqref{Eq:ephi}-\eqref{Eq:en}. 

In order to evaluate the BAU today following the above strategy, we notice that the mechanism is controlled by the following independent parameters ($i$) $y_{\phi ff}$ [inflaton-SM fermion effective coupling], ($ii$) $y_{\phi NN}$ [inflaton-RHN coupling], ($iii$) $M_1$ [the lightest RHN mass] and ($iv$) $\{Re[\theta_R],~Im[\theta_R]\}$ [constituents of $R$ matrix to estimate $Y_{\nu}$]. We will maintain a typical hierarchy of RHN masses as: $M_3 = 10^2 M_2 = 10^4 M_1$. Furthermore, the dynamics also depends on the input parameters from the inflaton potential: $\{n,\lambda\}$. As stated earlier, we consider $n=2$ in $V(\phi)$ of Eq. \eqref{eq:Vgen} and correspondingly the value of $\lambda= 2\times 10^{-11}$ is determined such that the inflationary observables like spectral index and tensor-to-scalar ratio are found to be within $95\%$ C.L. of the Planck+BICEP2/$Keck$  constraints. For a more detailed discussion, see the Appendix and ref. \cite{Garcia:2020eof,Martin:2010kz,Dai:2014jja,Ueno:2016dim}. Like the previous study \cite{Datta:2022jic}, here also we confine ourselves by choosing $y_{\phi ff}\lesssim \mathcal{O}(10^{-5}-10^{-4})$ \cite{Drewes:2017fmn,Garcia:2020eof} as above this value, non-perturbative production of fermions from inflaton decay may start to dominate. As a result, perturbative prescription for particle production would be invalid. \\

\begin{figure*}
	\centering
	\includegraphics[width=0.31\linewidth]{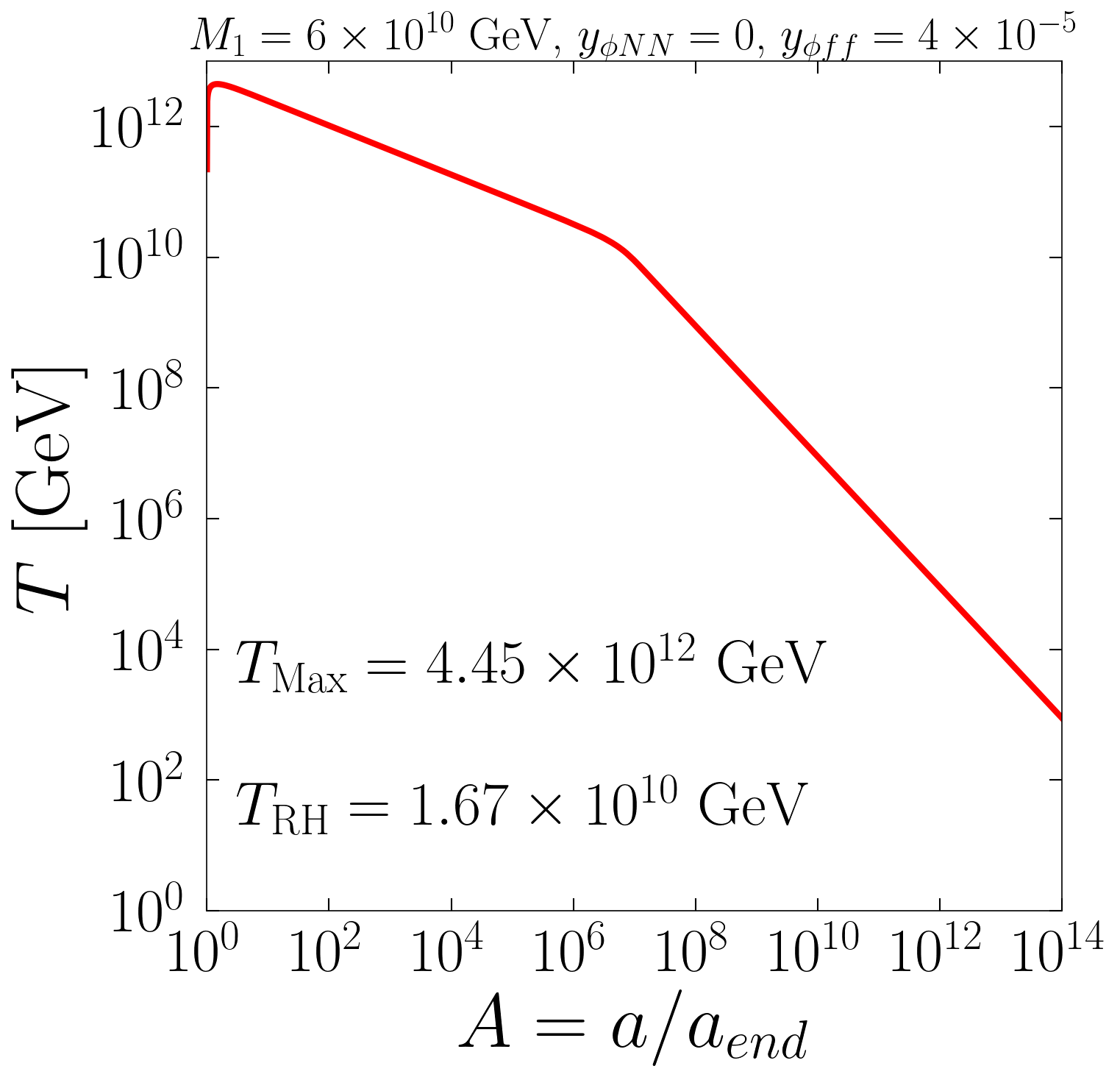}
	\includegraphics[width=0.31\linewidth]{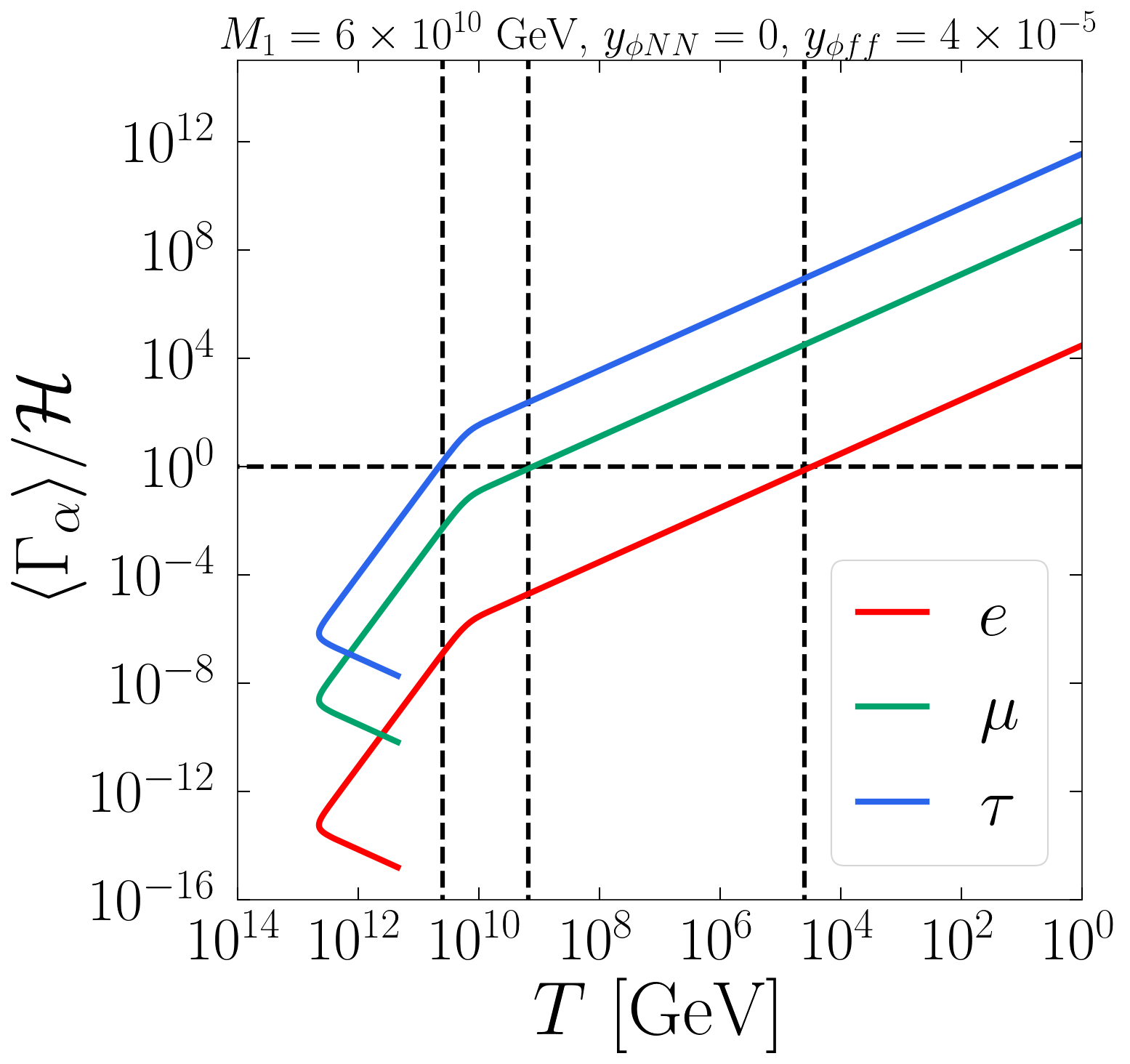}
	\includegraphics[width=0.32\linewidth]{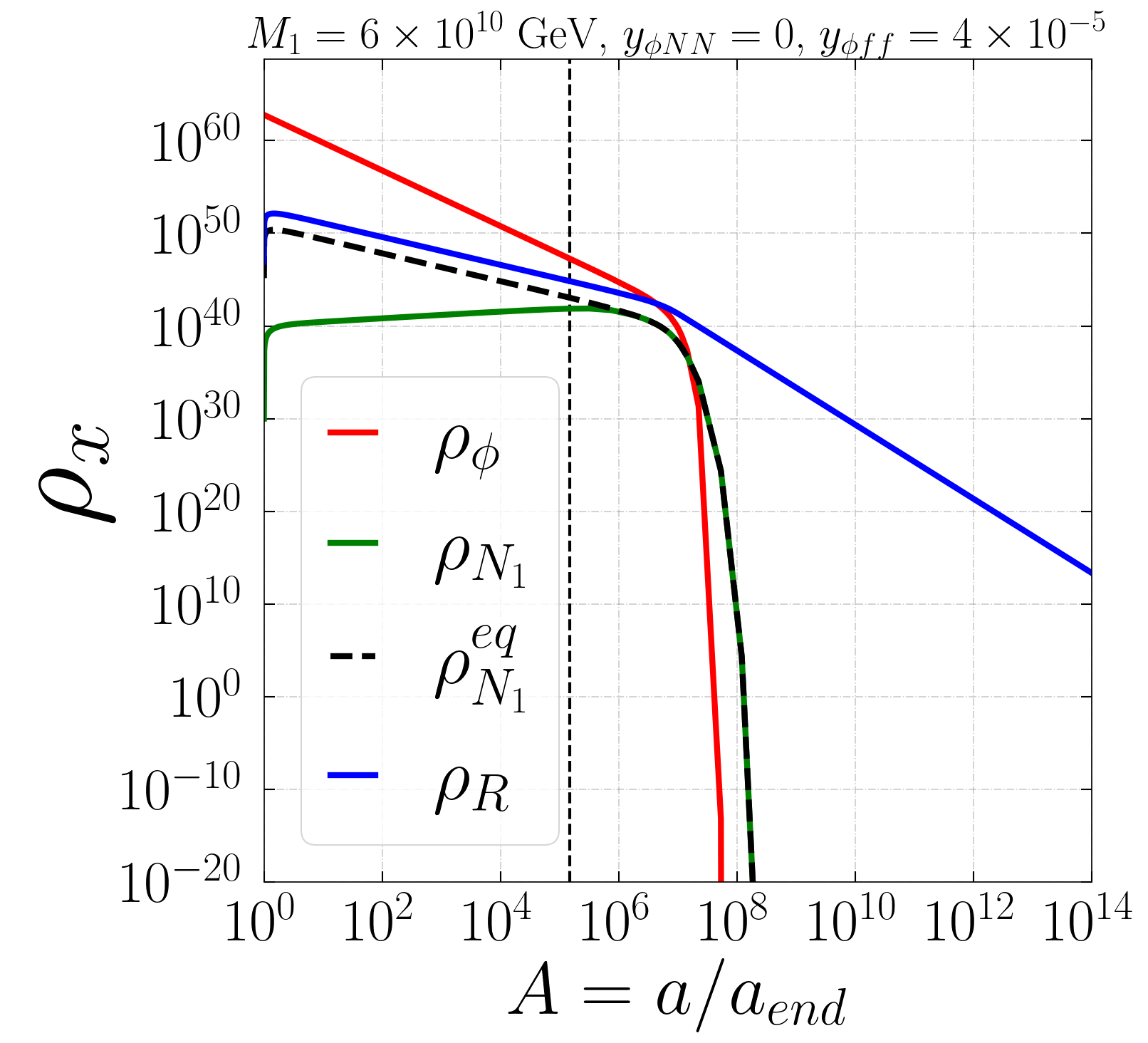}
	\captionsetup{justification=raggedright }
	\caption{ Evolution of temperature $T$ (left panel) and various energy densities (right panel) $w.r.t.$ rescaled scale factor for $M_1=6 \times 10^{10}$ GeV and $y_{\phi NN}=0$. In the middle plot we show the dependence of $\langle\Gamma_\alpha\rangle/\mathcal{H}$ on $T$ for the same choice of $M_1$ and $y_{\phi NN}$.}\label{fig:3}
\end{figure*}

\noindent {\bf{[In absence of inflaton-RHN coupling:]}} Note that the present scenario differs from the previous one due to the inclusion of $y_{\phi NN}$ coupling in this work. Hence, a choice of the parameter $y_{\phi NN} = 0$ should reproduce the outcome of our previous work. We therefore 
start studying the phenomenology by choosing $y_{\phi NN} = 0$  (case I) first and then turning on $y_{\phi NN}$ gradually to a value comparable to $y_{\phi ff}$. We choose $y_{\phi ff}= 4 \times 10^{-5}$ so as to be consistent with the perturbative limit on it. The Left plot of Fig.~\ref{fig:3} shows the variation of temperature (along $y$ axis) with the parametrized scale factor $A$ (along $x$ axis) where we use the solution for the radiation energy density of Eq. \eqref{Eq:r} (ignoring the first term in the right hand side as $y_{\phi NN} = 0$) coupled with Eqs. \eqref{Eq:ephi} and put it back in Eq. \eqref{eq:temp}. After inflation, the temperature of the Universe then attains a maximum value $\sim T_{\text{Max}}= 4.45 \times 10^{12}$ GeV and thereafter falls (having a different slope compared to the radiation dominated epoch) to a point where $\rho_{\phi} = \rho_R$ is reached which defined the end of reheating as $T_{\text{RH}}= 1.67 \times 10^{10}$ GeV. 

Due to such faster expansion and nontrivial scale factor dependence of temperature during this extended reheating period $T_{\text{Max}} > T > T_{\text{RH}}$, the charged lepton Yukawa interaction (particularly for $\tau_R$ in this case) will come to thermal equilibrium at a smaller temperature than the standard radiation-dominated case. To provide a concrete evaluation 
of the same, we include middle plot of Fig.~\ref{fig:3} where $\langle\Gamma_\alpha\rangle/\mathcal{H}$ evolution ($\alpha = e, \mu, \tau$  with blue, green and red lines respectively) are plotted against temperature $T$ variation. This shift in $\tau_R$ ET is depicted clearly by the intersection point of the blue line and the horizontal dashed line indicating $\langle\Gamma_\tau\rangle/\mathcal{H}=1$ in middle plot of Fig.~\ref{fig:3}. The relevant ET in this extended period of reheating turn out to be $T^*_{\tau}=4.7 \times 10^{10}$ GeV and is included in Table \ref{tab:table1} along with values of other parameters. The reheating temperature being bounded by $\sim \mathcal{O}(10^{10})$ GeV, no change in $\mu_R$ or $\tau_R$ ET has been found as expected. 

With such a shift in the ET of $\tau_R$ in this particular case (first row of Table \ref{tab:table1}), flavor leptogenesis would get affected. In order to have an impact of it on flavor leptogenesis, we choose a value of $N_1$ mass $M_1= 6\times 10^{10}$ GeV which falls intermediate between the associated $T_{\text{Max}}$ and $T_{\text{RH}}$. First we estimate the evolution of various energy densities $\rho_{\phi}, ~\rho_{R},~\rho_{N_1}$ in this scenario against $A$ by solving Eqs.~\eqref{Eq:ephi}-\eqref{Eq:en} simultaneously as shown in the rightmost plot of Fig.~\ref{fig:3} indicated by red, blue and green solid lines respectively. 
In evaluating $Y_{\nu}$, we consider $Re[\theta_R]=6.03,~Im[\theta_R]=0.22$ (the reason behind such a choice is to have correct BAU finally). Note that in absence of $y_{\phi NN}$ coupling, $N_1$s are thermally generated during $T_{\text{Max}} > T > M_1$ from the thermal bath consisting of SM fields, thanks to radiation production from inflaton decay via $y_{\phi ff}$. As the Universe expands, the radiation energy density is being diluted non trivially till an equality with inflaton energy density defining $T_{\text{RH}}$ (intersection of blue and red lines). Standard radiation domination follows only beyond this point. 

\begin{figure}[h]
	\centering
	\includegraphics[width=1\linewidth]{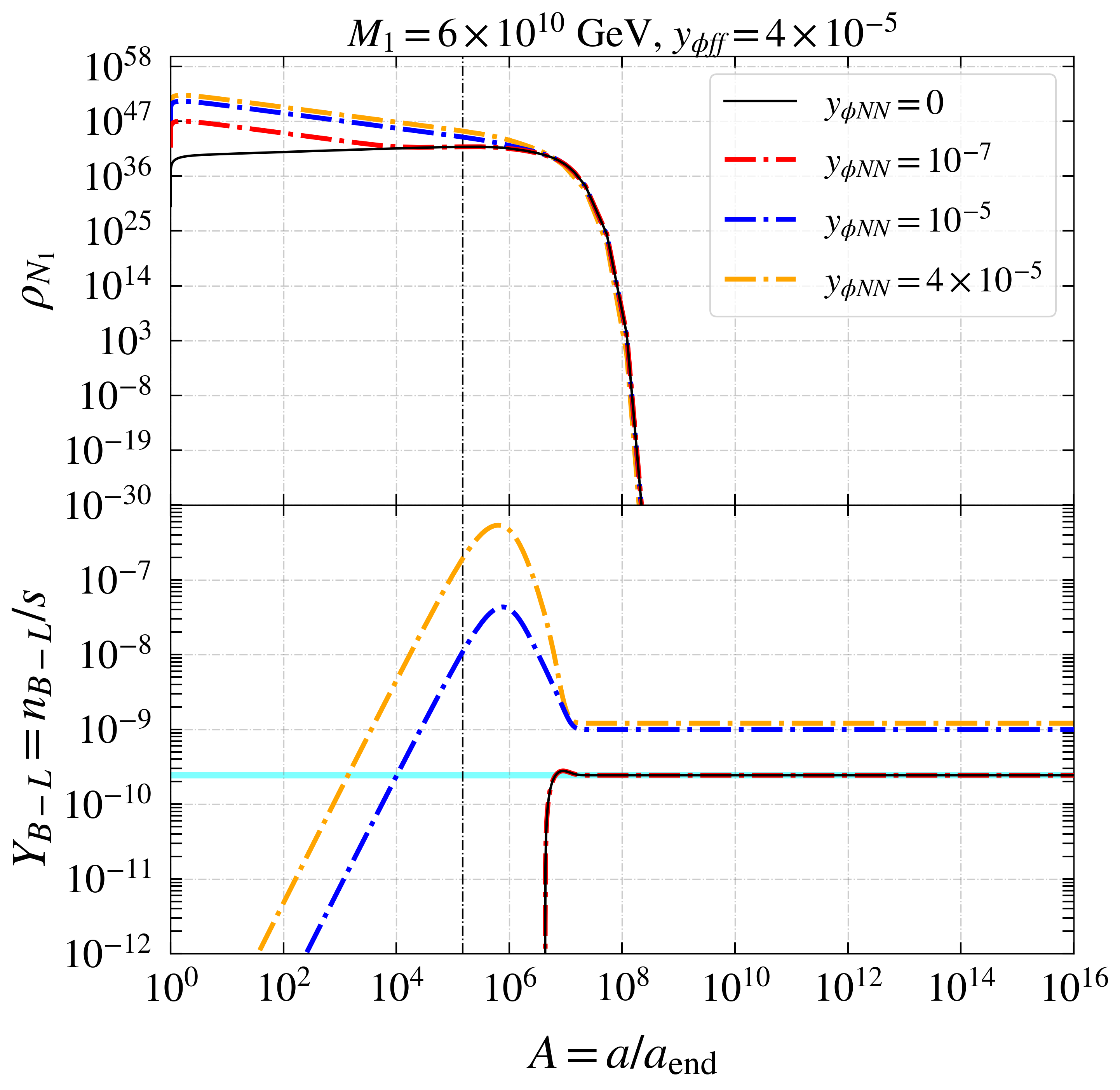}
	\captionsetup{justification=raggedright }
	\caption{ Evolution of energy density of $N_1$ (upper panel) and produced baryon asymmetry (lower panel) $w.r.t.$ rescaled scale factor for different values of $y_{\phi NN}$ for $M_1= 6\times 10^{10}$ GeV and $y_{\phi ff}=4 \times 10^{-5}$. }\label{fig:4}
\end{figure}

These $N_1$s would effectively decay around $T \sim M_1$ and produce the lepton asymmetry via leptogenesis. The $N_1$ being thermally produced, this scenario is similar to the standard flavored {\it thermal} leptogenesis scenario, though impacted by the shift in $T^*_{\tau}$ as seen above. We find $M_1 > T^*_{\tau}$, all the charged lepton Yukawa interactions remain out of equilibrium in this phase. As a consequence, an unflavored leptogenesis prevails here in case of extended period of reheating. This is the main difference we experience while comparing it with standard flavored {\it thermal} leptogenesis scenario in which with  leptogenesis scale $T \sim M_1$, $\tau$ lepton Yukawa interaction occurs rapidly (as it is already in equilibrium, see section~\ref{section:2}) following which a two-flavor setup must be incorporated (compared to the unflavored one in present case). 
The corresponding evolution of lepton asymmetry is shown by black line in bottom portion of Fig.~\ref{fig:4} as a function of the modified scale factor $A$ which saturates to a lepton asymmetry value that eventually converts to the observed BAU value. To make the correspondence between $\rho_{N_1}$ and the production of $Y_{B-L}$, we also incorporate the $\rho_{N_1}$ evolution in the top portion of the figure. \\

\begin{figure*}[t!]
	\centering
	\includegraphics[width=0.45\linewidth]{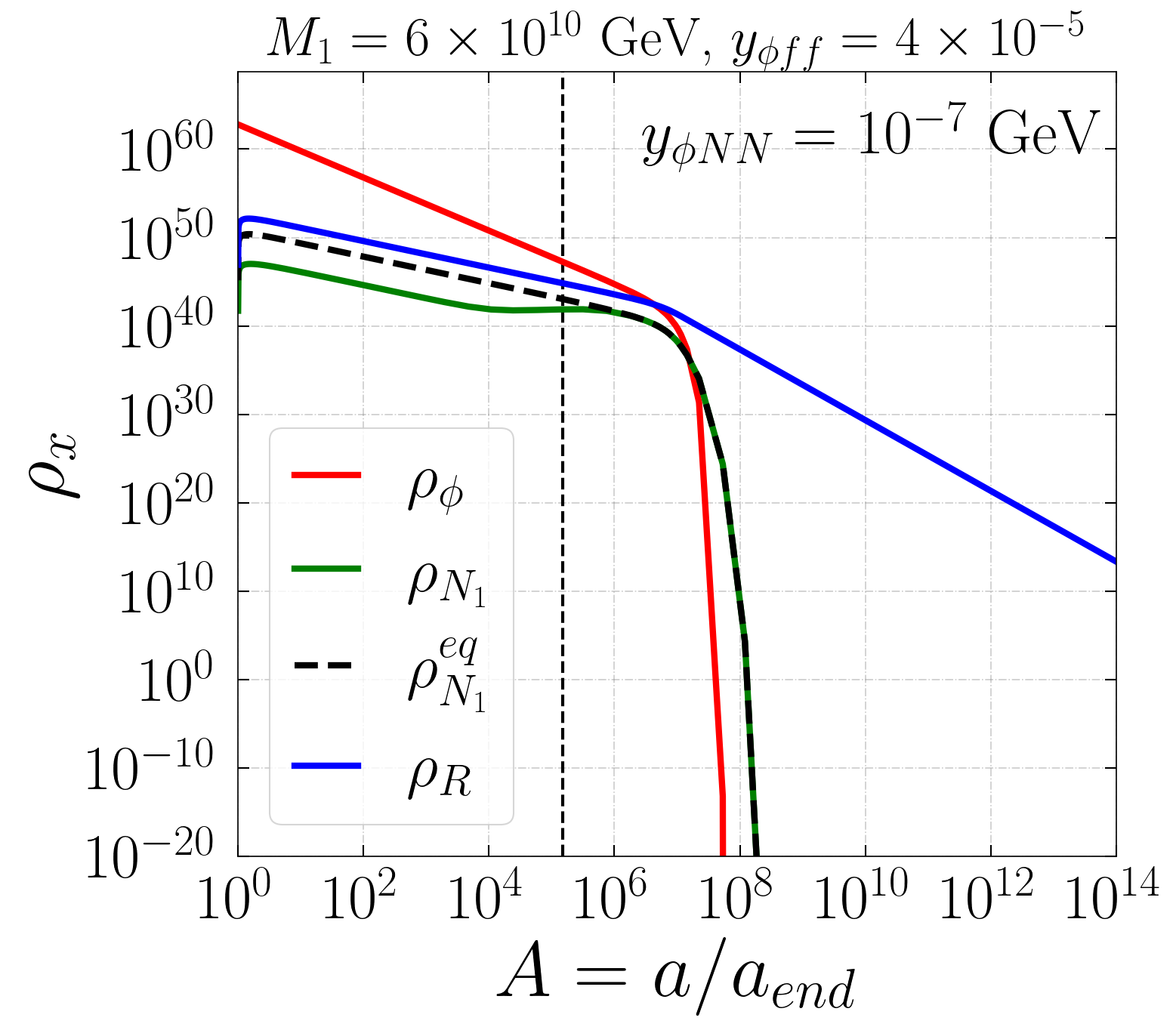}
	\includegraphics[width=0.45\linewidth]{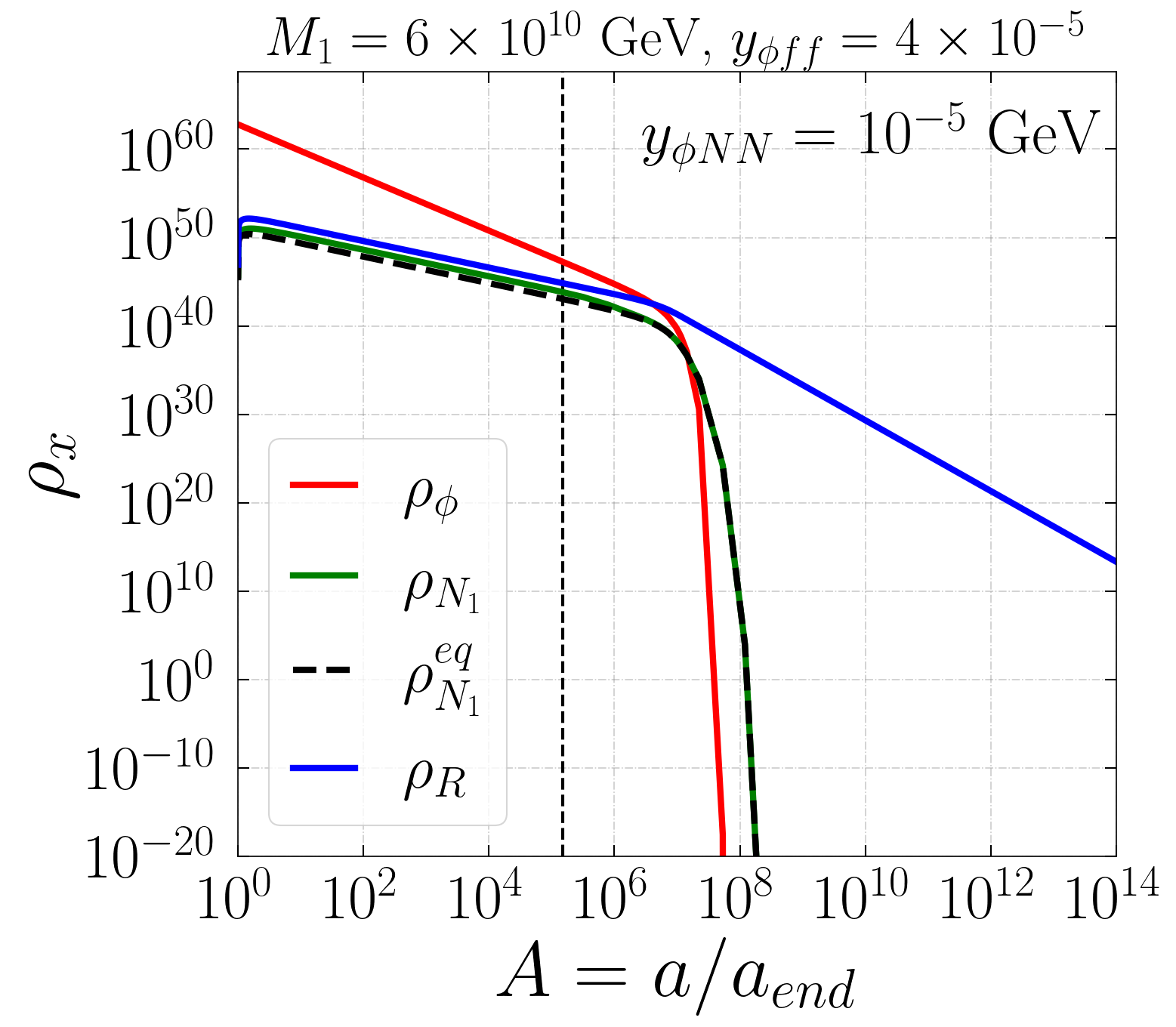}
	\captionsetup{justification=raggedright }
	\caption{ Evolution of different energy densities $w.r.t.$ rescaled scale factor for $y_{\phi NN}= 10^{-7}$(left panel) and $y_{\phi NN}= 10^{-5}$ (right panel). Here we set $M_1= 6\times 10^{10}$ GeV and $y_{\phi ff}= 4 \times 10^{-5}$. }\label{fig:5}
\end{figure*}

\noindent {\bf{[In presence of inflaton-RHN coupling:]}} We now turn on inflaton-RHN coupling and observe its impact on the charged lepton Yukawa equilibration and consequently on the produced baryon asymmetry during reheating era. Let us begin with a sufficiently small $y_{\phi NN} = 10^{-7}$ (tabulated in Table \ref{tab:table1} as case II) as compared to $y_{\phi ff}$ chosen. As shown in the Fig.~\ref{fig:5}, switching on $y_{\phi NN}$ causes the inflaton to produce a large number of $N_1$s (indicated by green line) initially. This can be understood if we compare the $\rho_{N_1}$ evolution (green line in Fig.~\ref{fig:5}) above temperature $T= M_1$ (indicated by the vertical black small dashed line) in this case versus the case with solely thermally produced $N_1$s (see Fig.~\ref{fig:3}). However, as the temperature drops, the production of $N_1$ from inflaton decay does not keep up with the Universe's expansion rate due to its feeble coupling chosen. As a result, the energy density of these $N_1$s (as decay products of inflaton) gets diluted and at some stage the production of $N_1$s from the inverse decay dominates over it. This is evident in the left plot of Fig.~\ref{fig:5} by the sudden change of slope of $\rho_{N_1}$ just before $T = M_1$ which coincides with the energy density of thermally produced $N_1$s of Fig.~\ref{fig:3}. This continues till a point beyond which $N_1$ decay starts to contribute to lepton asymmetry production.  

We also notice that due to the dominant decay of $\phi$ into SM fermions, the $\rho_{\phi}$ and $\rho_R$ (mainly contributed from $y_{\phi ff}$ coupling) do not alter by a noticeable amount and hence $\mathcal{H}$ remains essentially unchanged compared to the previous case with $y_{\phi NN}$ =0. As a result, together with $T_{\text{Max}}$ and $T_{\text{RH}}$, the $\tau_R$ ET $T^*_{\tau}$ remains identical with the {\it thermal} case (see second row of Table~\ref{tab:table1}). Hence the present situation falls in the category of unflavored leptogenesis. The evolution of the $B-L$ asymmetry for $y_{\phi NN}=10^{-7}$ scenario is presented by the red dash-dotted line in the bottom plot of Fig.~\ref{fig:4} which overlaps with the $y_{\phi NN}$ = 0 case (solid black line), thereby satisfying the observed baryon asymmetry of the Universe. 

As we further increase the strength of the inflaton-$N_1$ coupling, $i.e.$ $y_{\phi NN}=10^{-5}$ (as case III of Table \ref{tab:table1} while keeping other parameters fixed at their previous values), a change in $\rho_{N_1}$ becomes visible. The right plot of Fig.~\ref{fig:5} shows the behavior of the energy densities of different components of the Universe in this case. Note that contrary to cases I and II, dominant contribution to $\rho_{N_1}$ here follows from the $N_1$s being decay products of inflaton as it supersedes the thermally produced ones from inverse decay. The radiation component (blue line) however still remains dominant compared to $\rho_{N_1}$ (green line). Note that $T_{\text{Max}}$ remains unaffected as it is mainly controlled by $y_{\phi ff}$ coupling (fixed for cases I-IV) responsible for initial radiation production. However a small shift in the $T_{\text{RH}}$ as compared to cases I and II, is observed and indicated in Table \ref{tab:table1}. This happens as a result of higher $y_{\phi NN}$ coupling which causes the inflaton to decay earlier than cases I and II so that $\rho_{\phi} = \rho_R$ defining the $T_{\text{RH}}$ is realized at a slightly higher temperature. Due to the dominance of $\rho_R$ (almost unchanged compared case I and II) over $\rho_{N_1}$, the temperature evolution above $M_1$ remains close to the two earlier cases. For the same reason, $\mathcal{H}$ is also almost unaffected and this is reflected in the evaluation of $T^*_{\tau}$ (only a slight change) as included in Table \ref{tab:table1}. As previously discussed, here also, even though we have a slightly higher value of $T^*_{\tau}$ than the previous cases, the leptogenesis scale however remains larger with respect to $T^*_{\tau}$. Hence, an unflavored prescription is still adequate for estimating the lepton asymmetry.

The evolution of produced lepton asymmetry in this case III is shown by the blue dash-dotted line in the bottom panel of Fig.~\ref{fig:4}. As seen in this plot, the lepton asymmetry starts to being produced at a stage above temperature $T \sim M_1$. This is related to the fact that $N_1$s produced during this era of reheating find themselves in out of equilibrium ($\mathcal{H}$ is larger than decay width of $N_1$) and would decay. Additionally, $\rho_{N_1}$ being comparable to $\rho_{R}$, the inverse decay process (related with neutrino Yukawa coupling) remains subdominant compared to the $N_1$ decay. As a result, lepton asymmetry (still unflavored though) produced from such decay of $N_1$ would not be washed out completely and a non-zero asymmetry survives. The amount of asymmetry production increases till a point where $\rho_{N_1}$ ceases to exist. Beyond it, $Y_{B-L}$ falls to some extent, before attaining its asymptotic value which is larger than the value of the lepton asymmetry necessary for the production of observed BAU, as the produced asymmetry gets diluted due to the increase of entropy ($N_1$ decay produces a sizeable $\rho_R$ and hence, entropy) in the Universe. Note that this phase of leptogenesis is different from {\it thermal} leptogenesis scenario as $N_1$s are never in thermal equilibrium. On the other hand, this is not purely the case of {\it non-thermal} leptogenesis which happens with $N_1$, as the decay product of inflaton, finding itself in an environment with $T \ll M_1$ (so, thermal generation is ruled out). So here with case (B), we find a nonstandard generation of lepton asymmetry as a consequence of extended reheating where the inflaton has a sizeable coupling with the lightest RHN. As discussed in the beginning of section, we call it a `quasi'- thermal leptogenesis as neither it is the case of a purely thermal nor that of {\it non-thermal} leptogenesis. The value of $Y_{B}$ can be brought down to the correct BAU by decreasing $Im[\theta_R]=0.031$,  while all other parameters/outcomes are unaffected. 


Finally, the above discussed effect becomes prominent if we choose to increase the $y_{\phi NN}$ coupling further, say $y_{\phi NN} = y_{\phi ff}$ as included in case IV of Table~\ref{tab:table1}. In this case, we obtain $\rho_{N_1} = \rho_{R}$. As radiation and $N_1$s contribute equally to the energy density of the Universe during the reheating period, the expansion rate of the Universe gets modified in this scenario, affecting the $T_{\text{RH}}$ as well as $T^*_{\tau}$. In this case, we get a larger $T_{\text{RH}}$ as inflaton decays earlier than the previous case owing to the larger $y_{\phi NN}$ coupling. The related  numerical estimates for this case IV are listed in the fourth row of Table~\ref{tab:table1}. In this case also, a larger baryon asymmetry of the Universe is created which can be settled to the observed $Y_B$ value without altering any other parameter/predicted values once we reduce the value to $Im[\theta_R]=0.02$.

\begin{figure}[t!]
	\centering
	\includegraphics[width=1\linewidth]{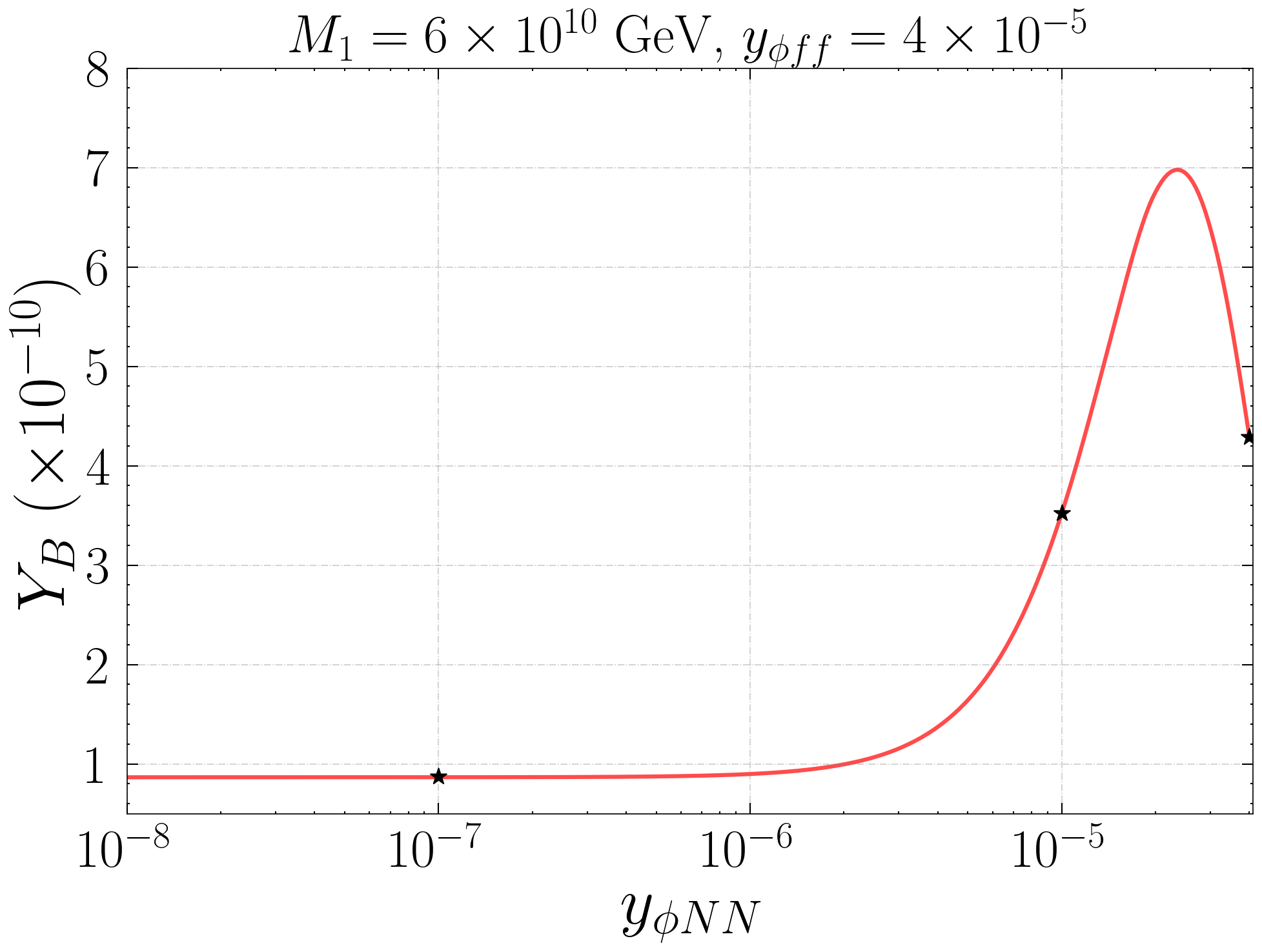}
	\captionsetup{justification=raggedright }
	\caption{Variation of final baryon asymmetry $Y_B$ $w.r.t.$ $y_{\phi NN}$ for $M_1= 6\times 10^{10}$ GeV and $y_{\phi ff}= 4 \times 10^{-5}$. The points indicated by ``star" represent the case-II-IV  from Table ~\ref{tab:table1}.}\label{fig:7}
\end{figure} 

\begin{table*}[t!]
	\begin{tabular}{||c|c| c|c| c| c|c|c|c||} 
		\hline
		Point &$y_{\phi NN}$ & $y_{\phi ff}$ &  $T_{\text{Max}}$ (GeV) & $M_1$ (GeV) &$T_{\text{RH}}$  (GeV)&  $T^*_{\tau}$ (GeV)& $T^*_\mu$ (GeV) & $Y_B$\\ [0.5ex] 
		\hline\hline
		BP1 & $10^{-6}$ & $10^{-7}$& $2.23 \times 10^{11}$&$5 \times 10^9$ & $4.2 \times 10^8$ &$2 \times 10^9$ & $7 \times 10^8$ & $5.67 \times 10^{-9}$\\
		\hline
		BP2 & $5 \times 10^{-7} $& $ 5 \times 10^{-8}$ & $1.6 \times 10^{11} $& $ 5 \times 10^9 $ & $ 2.1 \times 10^8 $& $1.5 \times 10^9$ &$4 \times 10^8$ & $3.60 \times 10^{-9}$\\
		\hline
		BP3 & $10^{-7}$ & $10^{-8}$& $7.04 \times 10^{10}$&$5 \times 10^9$ & $4.2 \times 10^7$ &$6 \times 10^8$ & $1.3 \times 10^8$ & $7.28 \times 10^{-10}$\\
		\hline
	\end{tabular}
	\captionsetup{justification=raggedright }
	\caption{\label{tab:table-nam} We list three benchmark points (BP) where the leptogenesis scale falls in between $T_{\text{Max}}$ and $T_{\text{RH}}$. While BP1 represents the case where $M_1$ is closer to $T_{\text{RH}}$, BP2 represents an intermediate scenario and BP3 indicates a scenario where $M_1$ lies closer to $T_{\text{Max}}$.}
\end{table*}

In Table \ref{tab:table1}, we have listed a few specific values of $y_{\phi NN}$ coupling to describe the impact of reheating on leptogenesis. In Fig.~\ref{fig:7}, we provide the estimate of final baryon asymmetry (via unflavored leptogenesis) once the Yukawa coupling $y_{\phi NN}$ is varied ($y_{\phi NN} \le y_{\phi ff}$). As already found in cases I-III, with tiny $y_{\phi NN}$ coupling ($y_{\phi NN} \lesssim 10^{-6}$), the final baryon asymmetry $Y_B$ almost remains independent of $y_{\phi NN}$. Thereafter, a rise in $Y_B$ can be seen due to the fact that the production of RHN $N_1$ from the inflaton decay also becomes significant. This additional production channel causes a significant rise in the $N_1$'s abundance $\rho_{N_1}$ which further leads to a larger production of lepton asymmetry (also the baryon asymmetry). This behavior is also clear from Fig.~\ref{fig:4}. A peak in $Y_B$ is observed when $y_{\phi NN} \simeq 2 \times 10^{-5}$ after which $Y_B$ is reduced once the $y_{\phi NN}$ is further increased. This fall can be understood by looking at the third term of Eq.~\eqref{eq:bau4} where one notes that a larger production of asymmetry also results in a larger washout of the asymmetry. \\

\noindent {\bf{[With dominant inflaton-RHN coupling:]}} So far the discussion we have, we find that the gradual increase of $y_{\phi NN}$ coupling not only affects the temperature behavior and expansion rate of the Universe during reheating period but also impacts the lepton asymmetry production in this {\it quasi}-thermal regime. However we have restricted ourselves with values of couplings associated to the inflaton below $\mathcal{O}(10^{-5}-10^{-4})$  so as to keep the analysis consistent with perturbative reheating era \cite{Drewes:2017fmn,Garcia:2020eof}. Alongside, we take $y_{\phi ff}$ at a borderline value $4 \times 10^{-5}$ and hence we are unable to make $y_{\phi NN}$ larger than $y_{\phi ff}$ by order(s) of magnitude and discuss the impact of such consideration. Also, with such choice of $y_{\phi ff}$, the reheating temperature turns out to be high enough so as to keep $M_1$ accordingly large (to realize scenario [B]). 

With an aim to observe the consequence of $y_{\phi NN} > y_{\phi ff}$ while keeping things more flexible such as lowering the scale of leptogenesis impacted by the extended period of reheating, we now consider three specific situations: (i) $M_1$ is close to $T_{\text{RH}}$ [BP1], (ii) $M_1$ is intermediate between $T_{\text{Max}}$ and $T_{\text{RH}}$ [BP2], and (iii) $M_1$ is close to $T_{\text{Max}}$ [BP3] where the mass of the lightest RHN is fixed at $M_1=5 \times 10^{9}$ GeV while $y_{\phi ff}$ and $y_{\phi NN}$ are floated to realize such considerations. We choose the values of $Re[\theta_R] = 2.83$ and $Im[\theta_R] = 0.24$ used in thermal flavored leptogenesis scenario (in two flavor regime) of section \ref{section:2} (see Fig.~\ref{fig:2}) the result of which is consistent with correct BAU. The purpose of such a choice is to compare the outcome of the extended period of reheating on final baryon asymmetry generation with $y_{\phi NN} > y_{\phi ff}$.
\begin{figure*}
	\centering
	\includegraphics[width=0.31\linewidth]{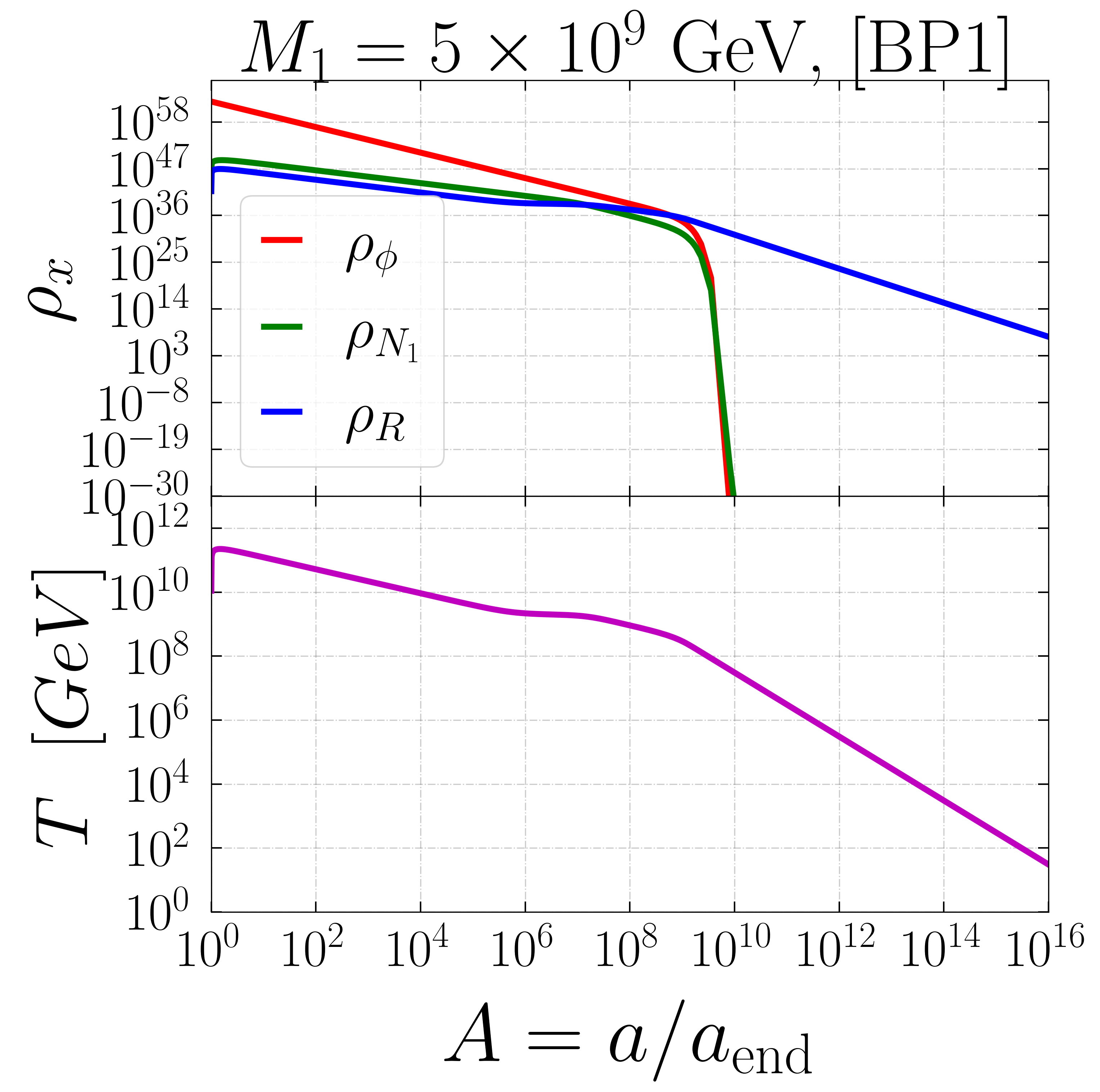}
	\includegraphics[width=0.32\linewidth]{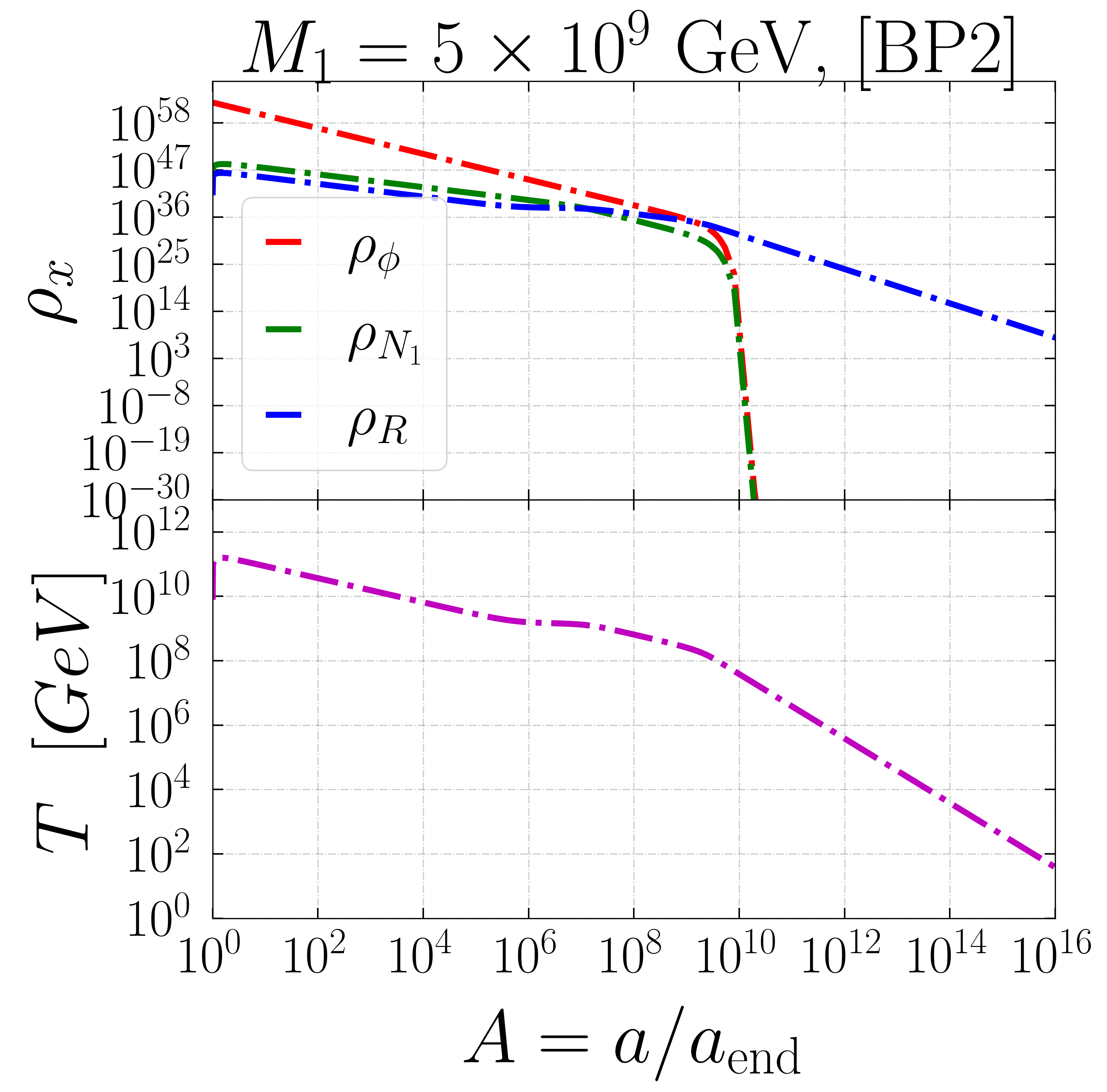}
	\includegraphics[width=0.31\linewidth]{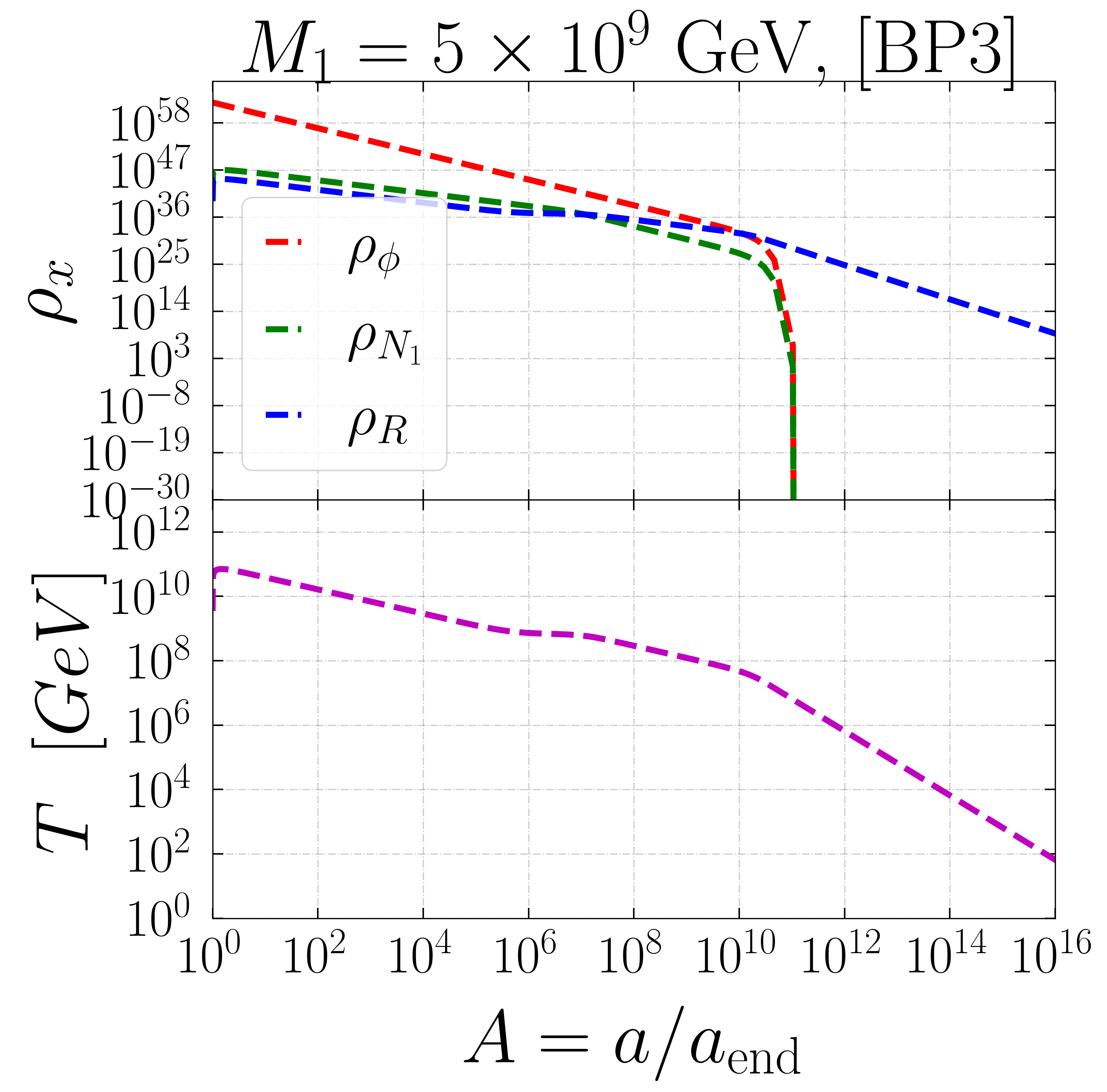}
	\captionsetup{justification=raggedright }
	\caption{ Evolution of different energy densities (upper panel) and temperature $T$ (lower panel) $w.r.t.$ rescaled scale factor for BP1 (left panel), BP2 (middle panel), and BP3 (right panel). Here we fix $M_1= 5\times 10^9$ GeV and $\theta_R= 2.83+0.24 i$.}\label{fig:8}
\end{figure*}
\begin{figure}
	\includegraphics[width=1\linewidth]{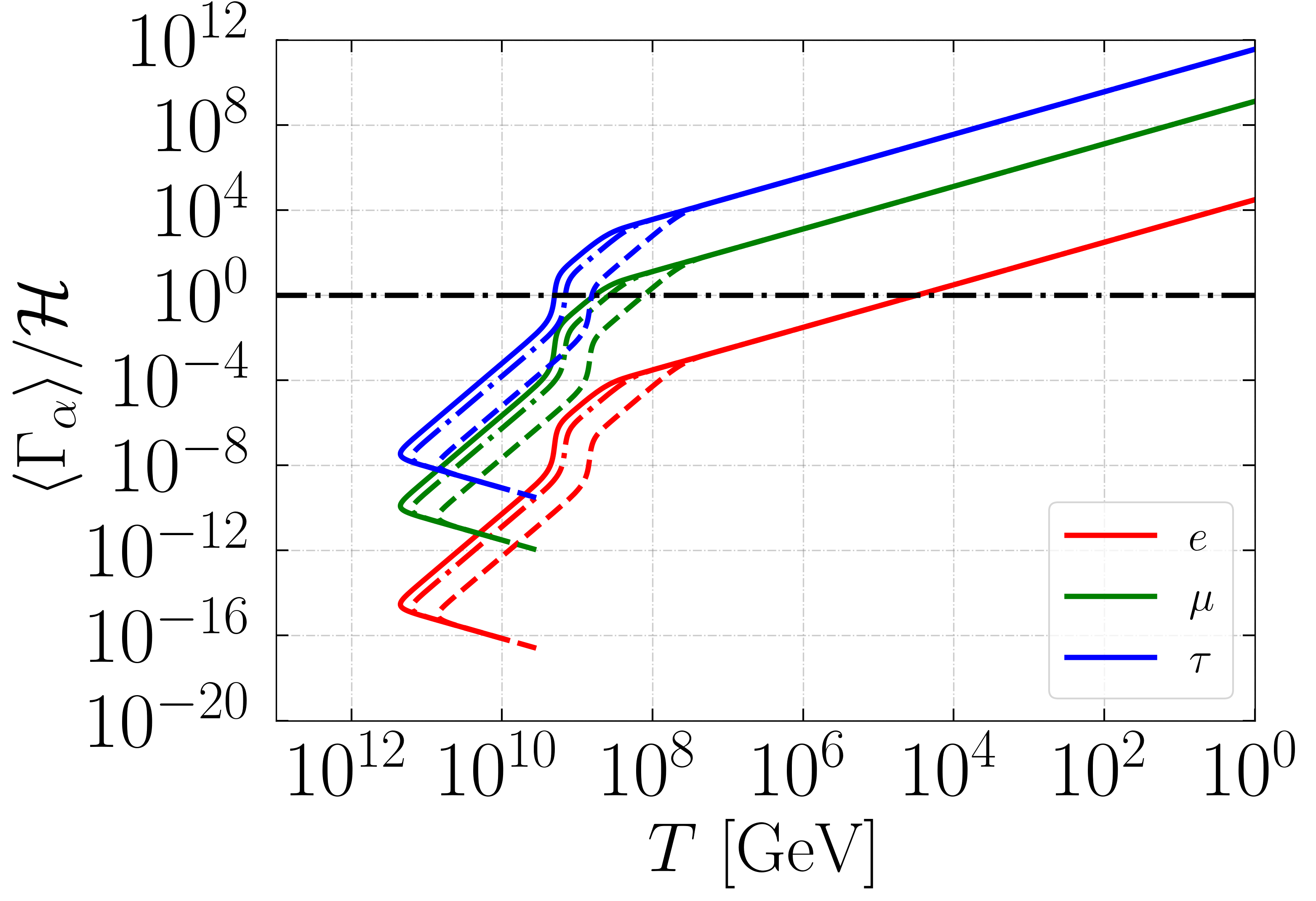}
	\captionsetup{justification=raggedright }
	\caption{ Variation of $\langle\Gamma_\alpha\rangle/\mathcal{H}$ $w.r.t.$ $T$ for  $M_1= 5\times 10^9$ GeV and $\theta_R= 2.83+0.24 i$. Here solid lines indicate the BP1, dashed dotted line represents BP2, and BP3 is denoted by dashed lines.  }\label{fig:9}
\end{figure}
\begin{figure}
	\includegraphics[width=1\linewidth]{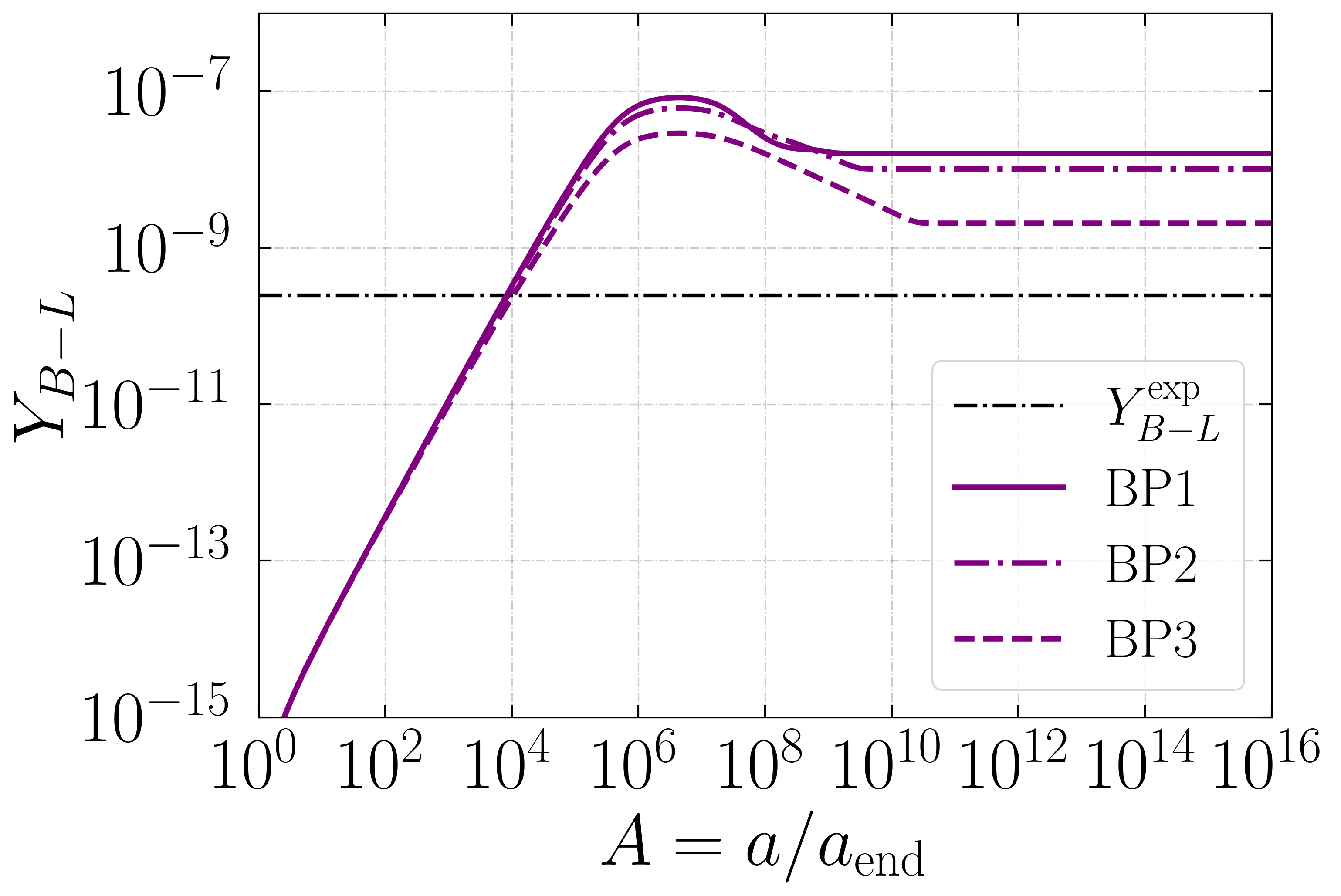}
	\captionsetup{justification=raggedright }
	\caption{ Evolution of produced lepton asymmetry $w.r.t.$ rescaled scale factor for all the three BPs for  $M_1= 5\times 10^9$ GeV and $\theta_R= 2.83+0.24 i$.}\label{fig:10}
\end{figure}

Note that as inflaton-SM fermion coupling essentially define the $T_{\text{Max}}$ while $y_{\phi NN}$ has some role to play in determining $T_{\text{RH}}$, we first make some appropriate choices of these two parameters in defining the three benchmark cases BP-1,2,3. They are listed in Table~\ref{tab:table-nam}. For all these sets, $y_{\phi NN}$ remains one order of magnitude larger than $y_{\phi ff}$ coupling. In evaluating the temperature evolution during the reheating, 
we solve Eqs.~\eqref{Eq:ephi}-\eqref{Eq:en} as a function of the rescaled scale factor $A$ simultaneously and using Eq.~\eqref{eq:temp}, temperature is evaluated. Bottom panels of all three plots of Fig.~\ref{fig:8}  represent the temperature variation with $A$ and upper panels depict the same for energy densities of different components. 

As seen from the plots, immediately after the end of inflation, the temperature reaches a maximum value $T_{\text{Max}}$. Then 
it starts to decrease in accordance with our previous discussion (in section \ref{section:4}) due to faster expansion of the Universe during this period of extended reheating. However, an interesting departure of $T$ from this fall is observed around $T \sim M_1$. This is related to the emergence of a new production channel, producing radiation from $N_1$ decay asa result of $y_{\phi NN}$ dominance. An interplay between such additional injection of radiation in the bath (which tries to increase 
$\rho_R$) and the depletion of $\rho_R$ due to Hubble expansion, a plateau like region is formed in $T$ evolution plot. 
This period however does not last long as eventually Hubble expansion rate overtakes this radiation production rate from $N_1$ decay ($\rho_{N_1}$ decreases sharply beyond a point). Eventually, radiation dominates over matter beyond $T_{\text{RH}}$ and temperature of the Universe drops as $A^{-1}$. In the upper panels of Fig.~\ref{fig:8}, we note that due to choice(s) of smaller coupling(s) $y_{\phi ff}$ ($y_{\phi NN}$) in going from BP1 to BP3, inflaton takes larger time to decay for BP3 (BP2)compared to BP1. As a result, matter radiation equality shifts at a later epoch resulting in lower reheating temperature for BP3 (BP2) relative to BP1.

This nontrivial behavior of temperature along with the larger expansion rate of the Universe during reheating period affects the ETs of charged lepton Yukawa interactions as can be seen from Fig.~\ref{fig:9}. For RHN mass $M_1= 5\times 10^9$ GeV, though there is no shift in ET for right-handed electron, ample amount of change can be seen for ET for $\mu_R$ and $\tau_R$ compared to the standard radiation dominated scenario for all three BPs. This change makes the charged Yukawa interactions come to equilibrium at a much lower temperature which are included in Table \ref{tab:table-nam}.  

As a consequence of considerably lower values of $T^*_{\mu}$ and $T^*_{\tau}$, we expect the quantum decoherence of the SM lepton doublet states to take place here at much lower temperatures (same as charged lepton ET) compared to the standard radiation dominated scenario. Hence, for all three BPs, lepton asymmetry generation process turns out to be not affected by individual charged lepton doublets at leptogenesis scale $M_1= 5\times 10^9$ GeV and an unflavored leptogenesis prevails here. This is a new result as compared with earlier standard analysis presented in section \ref{section:2}, where at this leptogenesis scale, $\tau_R$ was already in equilibrium affecting the lepton asymmetry generation along $\tau$ direction distinctively. Accordingly, a two flavor leptogenesis was incorporated for correct generation of lepton asymmetry. On the other hand, in this present case incorporating the effect extended reheating (with inflaton-RHN dominance), unflavor approach to evaluate the baryon asymmetry of the Universe would be enough.

For all three BPs, a different rate of washout during the reheating period accounts for the main difference in the produced final baryon asymmetry. Dashed purple line of Fig.~\ref{fig:10} shows that as $M_1$ is closer to $T_{\text{Max}}$ for BP3 as a result of which the produced asymmetry suffers a larger amount of washout (due to Hubble expansion) contrary to BP1 and BP2. Finally, even with a relatively low $M_1$ in this {\it quasi}-thermal regime, an overproduction of baryon asymmetry by one to two order(s) of magnitude is observed for these three BPs relaxing the parameter space even further with respect to the modified thermal leptogenesis scenario studied in section \ref{section:5}. For completeness purpose, we notice that such final $Y_B$ values can be brought down to correct level of BAU by changing $\theta_R$ without altering other parameter values or the outcomes such as $T_{\text{RH}}$ and $T^*_{\alpha}$. For example, for BP1 (BP2), one needs to set $Im[\theta_R]= 0.002~(0.005)$, while for BP3,  $Im[\theta_R]$ can be fixed at $0.02$ so that observed BAU can be generated.

\section{Conclusion}
\label{conclusion}

In this work, we have shown that an extended period of reheating resulting from inflaton-decay into radiation  
together with the lightest RHN can significantly alter the equilibration temperature of the charged lepton Yukawa interactions. Consequently, flavored leptogenesis mechanism gets affected. We start with a discussion on how the equilibration temperature(s) of charged lepton Yukawa interaction(s) can be estimated in a radiation dominated Universe and its impact on lepton asymmetry generation known as flavored leptogenesis. In such a setup, the reheating process is generally assumed to be instantaneous and happens to be higher than the mass of the decaying RHN whose decay contributes to lepton asymmetry production. However, depending on the inflaton coupling to SM particles, the reheating process may survive a longer period creating a prolonged era of reheating, from $T_{\text{Max}}$ to $T_{\text{RH}}$. Motivated by our recent finding on the impact of this extended era of reheating on charged lepton equilibration temperature and flavored leptogenesis, here we extend the setup by including additional inflaton-RHN coupling. 

We find that with relatively large value of inflaton-RHN coupling compared to the inflaton-SM fermion effective coupling, the reheating period gets further modified. While the inflaton-SM fermion coupling mainly controls the maximum temperature of the Universe immediately after inflation, the inflaton-RHN coupling has the potential to impact the reheating temperature. 
The production of RHN and SM bath from the inflaton decay during this period of prolonged reheating helps the Universe to expand at a much faster rate (depending on the inflaton-RHN coupling though) in comparison to the scenario where inflaton decays directly to radiation solely. As a result of such faster expansion, along with the modified temperature behavior, the charged Yukawa interactions enter into equilibrium in a delayed fashion. We also observe that such a delayed equilibration of charged lepton Yukawa interactions can significantly modify the lepton asymmetry generation compared to what is observed in {\it thermal} leptogenesis. For example, a flavored leptogenesis scenario found to be in two flavor regime in standard {\it thermal} leptogenesis may emerge as an unflavored one here. 

Another interesting outcome of the present scenario is revealed with a dominant inflaton-RHN coupling with respect to inflaton-SM fermion effective coupling. Here we encounter an unusual situation where the lepton asymmetry starts to be produced at a temperature above the mass of the lightest RHN without being completely washed out. In fact, the reheating era produces an environment where the lightest RHN find itself in out-of-equilibrium in this regime and its decay therefore contributes to lepton asymmetry production. In a way, this helps to reduce the scale of leptogenesis since the inclusion of inflaton-RHN coupling may inject a large amount of RHN into the system on top of thermally produced ones (whose decay also contribute to produce lepton asymmetry) during reheating thereby resulting an enhanced lepton asymmetry. 
The framework however can be extended beyond our present consideration. In some of the low scale leptogenesis scenarios, the framework might alter the prediction as well as allowing the leptogenesis scale to drop even further opening the possibility to explore leptogenesis in collider experiments. The related study is beyond the scope of this paper and we plan for some more work in these directions in future. 

\section{Acknowledgements}
A.D. acknowledges the financial support received from the grant CRG/2021/005080.
A.S. acknowledges the support from grants CRG/2021/005080 and MTR/2021/000774 from SERB, Govt. of India. R.R. acknowledges the National Research Foundation of Korea (NRF) grant funded by Korea
government (NRF-2020R1C1C1012452).

\appendix

\section{Predictions of inflatonary observables}
The inflation potential in our work follows from supergravity framework \cite{Kallosh:2013hoa} and is of the form 
\begin{align}
	V(\phi)= \lambda M^4_P \left[\sqrt{6} \tanh\left(\frac{\phi}{\sqrt{6}M_P}\right)\right]^n. 
	\label{apeq:pot}
\end{align}
It contains free parameters $\lambda$ and $n$. In order to estimate the inflationary predictions followed from such a potential, the slow-roll parameters need to be considered, which are given by:
\begin{align}
	\epsilon=\frac{1}{2}M_P^2 \left(\frac{V'}{V}\right)^2, \quad
	\eta=M_P^2 \left(\frac{V''}{V}\right).
\end{align}
On the other hand, the number of $e$-folds can be estimated as:
\begin{align}
	N&= \frac{1}{M_P^2} \int_{\phi_{\rm end}}^{\phi_*} \frac{V}{V'}d\phi\simeq \int_{\phi_{\rm end}}^{\phi_*}\frac{1}{\sqrt{2 \epsilon}M_P}d\phi, \notag\\
	&\simeq \frac{3}{2 n}\cosh\left(\sqrt{\frac23}\frac{\phi_*}{M_P}\right),
	\label{apeq:nstar1}
\end{align}
where $\phi_*$ corresponds to the crossing horizon value of the inflaton and $\phi_{\rm end}$ represents the field value at the end of inflation. 

Consequently, within the slow-roll approximation, the inflationary observables such as spectral index ($n_s$) and the tensor-to-scalar ratio ($r$) can be expressed as:
\begin{align}
	n_s&\simeq 1-6 \epsilon +2 \eta, 
	\label{apeq:ns}	\\
	r&\simeq16 \epsilon = \frac{12}{N^2}.
	\label{apeq:r}
\end{align}
Finally, the remaining inflationary observable, the amplitude of the curvature power spectrum $P_{s}$, is given by:
\begin{align}
	P_{s} = \frac{V^3}{12 \pi^2 M_P^6 V'^2}.
	\label{apeq:as1}
\end{align}
All these three inflationary observables are to be evaluated at $\phi = \phi_*$. 

Note that contrary to two other parameters, $P_s$ is a function of normalization constant $\lambda$ (involved in the potential of Eq.~\eqref{apeq:pot}). 
Therefore, using the relation between $\phi_*$ and $N$ from Eq.~\eqref{apeq:nstar1}, $P_{s}$ at $\phi_*$ ($\equiv P_{s*}$) can be simplified as:
\begin{align}
	P_{s*}\simeq \frac{6^{n/2}\lambda N^2}{18 \pi^2},
	\label{apeq:as2}
\end{align}
which helps to determine the value of $\lambda$ for a specific $n$ since at the {\it Planck} pivot scale, $k_*=0.05$ Mpc$^{-1}$, $\ln(10^{10} P_{s*}) = 3.044$ holds \cite{Planck:2018vyg}. 
A precise estimate of the number of $e$-foldings $N$ would be helpful in determining these observables. In general, $N$ can be influenced by the duration of the reheating process which turns out to be important to consider here as our analysis is intricately connected to the reheating temperature. A better estimate of $N$ (considering no additional entropy production between the end of reheating $T_{RH}$ and the re-entry of the pivot scale $k_*$ to the horizon at a later epoch) follows as~\cite{Martin:2010kz, Liddle:2003as}
\begin{widetext}
	\begin{align}
		N = \frac14 \ln\left[\frac{V(\phi_*)^2}{M_P^4 \rho_{\rm end}}\right]+ \frac{1-3 \langle\omega\rangle}{12 (1+\langle \omega \rangle)}\ln \left[\frac{\rho_{\rm RH}}{\rho_{\rm end}}\right]
		+\ln\left[\frac{1}{\sqrt{3}}\left(\frac{\pi^2}{30}\right)^{1/4}\left(\frac{43}{11}\right)^{1/3}\frac{T_0}{H_0}\right]
		-\ln\left[\frac{k_*}{a_0 H_0}\right]-\frac{1}{12}\ln g_{\rm RH},
		\label{apeq:nstar2}
	\end{align}
\end{widetext}
where $\langle \omega \rangle$ is the $e$-fold average of the equation-of-state parameter $\omega$ during reheating and $g_{\rm RH}$ denotes the effective relativistic degrees-of-freedom at reheating, taken as the SM value = $106$. The present day temperature and Hubble parameter are $T_0=2.7255$ K and $H_0=67.36$ km s$^{-1}$ Mpc$^{-1}$ respectively following the CMB observation \cite{Planck:2018vyg}. The present-day scale factor $a_0$ can be set to unity without loss of any generality. Using Eqs.~\eqref{apeq:as1}-\eqref{apeq:as2} and an approximated $N$ expression: $N \simeq \sqrt{3/2} V (\phi_*)/[M_P V'(\phi_*)]$, the value of $V(\phi_*)$ can be estimated as $V(\phi_*) \simeq 6^{n/2}\lambda M_P^4$. The energy density at the end of inflation, $\rho_{\rm end}$, can be evaluated using $\phi_{\rm end}$ obtained by setting the slow roll parameters to be one. On the other hand, $\rho_{\rm RH}$ corresponds to energy density of radiation at $T_{\rm{RH}}$. Finally, $\langle \omega \rangle$ is determined by identifying it for the $\phi$ field only ($\langle\omega_\phi\rangle$) and 
that too is estimated over a complete cycle of inflaton oscillation about origin in post-inflationary reheating phase, which is found to be~\cite{Garcia:2020eof} $\langle\omega_\phi\rangle=\frac{n-2}{n+2}$. 

\begin{figure}[t!]
	\includegraphics[scale=0.25]{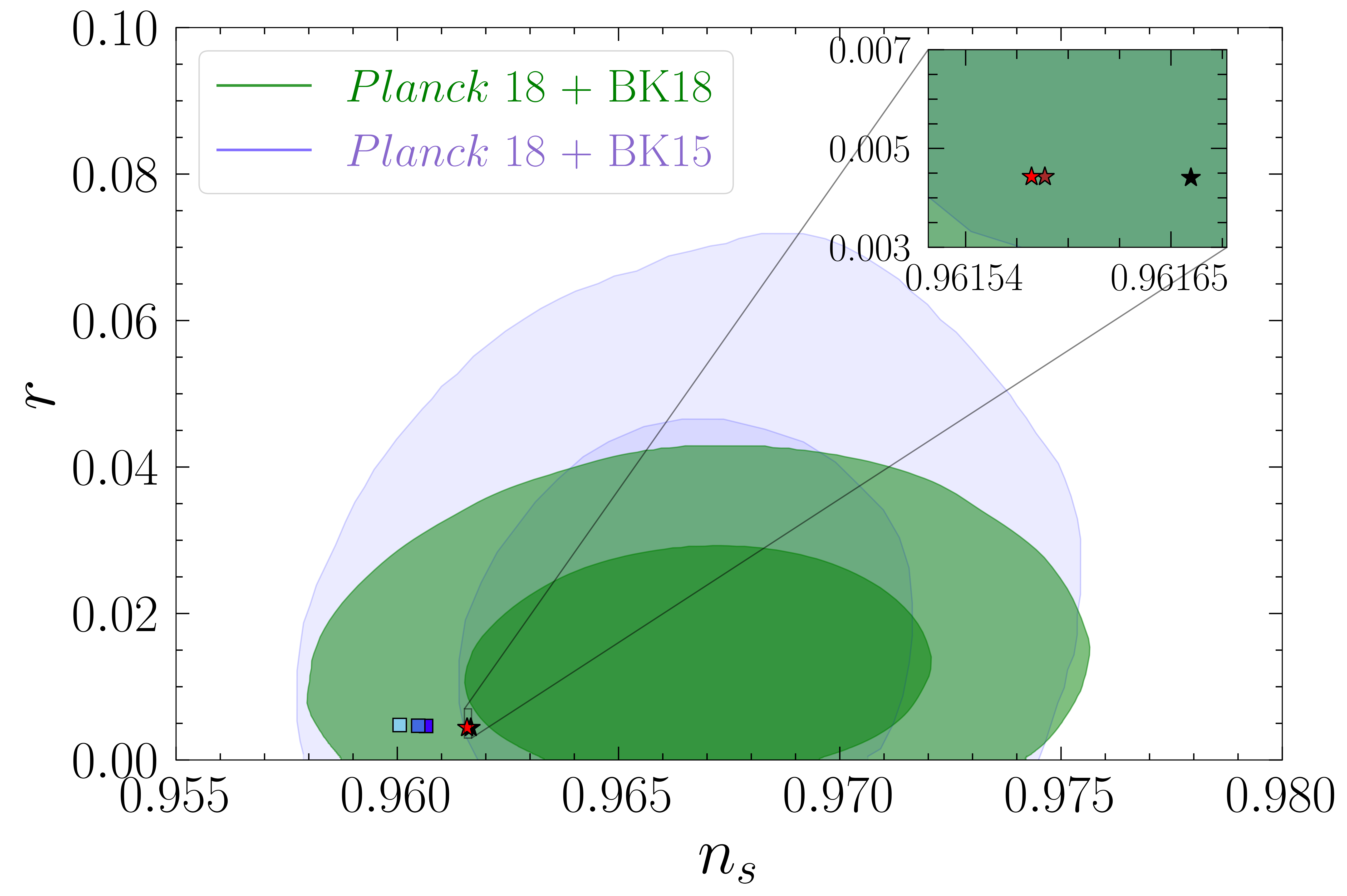}
	\caption{$68\%$ (light patch) and $95\%$ (dark patch) contours in the  $r$ vs. $n_s$ plane with 
		$Planck$ 2018\cite{Planck:2018vyg} + BICEP$/Keck$ (BK) 2015 data\cite{BICEP2:2018kqh} [$Planck$ 2018 + BICEP$/Keck$ 2018\cite{BICEP:2021xfz} data]:  drawn in purple [green] patches. 
		Red and brown stars are the ($n_s, r$) values generated from Table~\ref{tab:table1} while BPs from Table~\ref{tab:table-nam} produce the blue blocks. For a better view, the inset plot provides a zoomed view of the calculated $(n_s, r)$ from Table~\ref{tab:table1}. \label{apfig:a}}
\end{figure}

We restrict ourselves with $n = 2$ in the study. However the analysis can easily be extended for a more general set-up with the discussion above.  Now, with $n = 2$, we use different values of $\rho_{\rm RH}$ from Table~\ref{tab:table1} and \ref{tab:table-nam} of main text and proceed to solve Eq.~\eqref{apeq:nstar2} numerically to finally determine the set of $n_s$ and $r$ parameters (from each values of $T_{\rm{RH}}$ values) using Eqs.~\eqref{apeq:ns} and \eqref{apeq:r} and an estimate of $\lambda$ follows as $2 \times 10^{-11}$. We find $\phi_{\rm end} \simeq M_P$. The evaluated values of $n_s$ and $r$ are plotted in Fig.~\ref{apfig:a}, where red, brown and black stars indicate the set of $(n_s,r)$ values generated from cases I, III and IV of Table~\ref{tab:table1} respectively. On the other hand, BP1, BP2 and BP3 from Table~\ref{tab:table-nam} produce the $(n_s, r)$ values represented by dark blue, blue and light blue blocks respectively. The light and dark purple (green) patches of the figure represent the allowed range of $(n_s, r)$ obtained from the combined {\it Planck} 2018\cite{Planck:2018vyg} and BICEP/$Keck$ 2015 analysis \cite{BICEP2:2018kqh} (combined {\it Planck} 2018 and BICEP/$Keck$ 2018 analysis \cite{BICEP:2021xfz}) at  $68\%$ confidence level (C.L.) and $95\%$ C.L. respectively. As can be seen from the figure, cases discussed in Table~\ref{tab:table1} produce  $(n_s, r)$ values which are within the $95\%$ C.L. (or close to $95\%$ C.L. with updated analysis). On the other hand, the $(n_s, r)$ values produced from the BP1, BP2 and BP3 of Table~\ref{tab:table-nam} fall within the $68\%$ C.L. \\

\bibliography{ref.bib}
\end{document}